\documentclass[review,hidelinks,11pt]{elsarticle}
\usepackage[left=1.5cm,right=1.5cm, top=1.5cm, bottom=1.5cm]{geometry}
\usepackage{url}
\usepackage{setspace} 
\usepackage{lmodern}
\usepackage{graphicx}
\usepackage{color}
\usepackage{float}
\usepackage{comment}
\usepackage{subcaption}
\usepackage{amsmath}
\usepackage{wasysym}
\usepackage{booktabs}
\usepackage[colorlinks=true, citecolor=blue, linkcolor=blue, urlcolor=red]{hyperref}
\usepackage[figurename=Fig.]{caption}
\usepackage[tablename=Table]{caption}
\usepackage{tikz}
\newcommand{\tikzcircle}[2][red,fill=red]{\tikz[baseline=-0.75ex]\draw[#1,radius=#2] (0,0) circle ;}%
\newcommand{\tikzsquare}[2][red,fill=red]{\tikz[baseline=0ex]\draw[#1,#1] (0,0) rectangle (0.25cm,0.25cm); }
\newcommand{\tikztriangle}[2][red,fill=red]{\tikz[baseline=0ex]{\draw[#1,#1] (0,0) --
(0.25cm,0) -- (0.125cm,0.25cm) -- (0,0);} }
\newcommand{\tikzdiamond}[2][red,fill=red]{\tikz[baseline=-0.75ex]{\draw[#1,#1] (0,-0.15cm) --
(0.1cm,0) -- (0,0.15cm) -- (-0.1cm,0) -- (0,-0.15cm);} }
\newcommand{\tikzplus}[2][red,fill=red]{\tikz[baseline=-0.75ex]{\draw[#1,#1] (0,0) -- (-0.15cm,0) -- (0,0) -- (0.15cm,0) -- (0,0) -- (0,0.15cm) -- (0,0) -- (0,-0.15cm) -- (0,0);} }
\newcommand{\tikzcross}[2][red,fill=red]{\tikz[baseline=-0.75ex]{\draw[#1,#1] (0,0) -- (-0.125cm,-0.125cm) -- (0,0) -- (0.125cm,0.125cm) -- (0,0) -- (-0.125cm,0.125cm) -- (0,0) -- (0.125cm,-0.125cm) -- (0,0);} }
\newcommand{\tikzline}[2][red,fill=red]{\tikz[baseline=-0.75ex]{\draw[#1,#1] (-0.2cm,0) -- (0,0) -- (0.2cm,0);}}

\journal{Journal of Non-Newtonian Fluid Mechanics}

\begin{document}

\begin{frontmatter}

\title{Hydrodynamic interaction of a bubble pair in viscoelastic shear-thinning fluids}

\author[mymainaddress]{Mithun Ravisankar}
\author[otheraddress]{Alam Garcidueñas Correa}
\author[mymainaddress]{Yunxing Su}
\author[mymainaddress]{Roberto Zenit \corref{mycorrespondingauthor}}
\cortext[mycorrespondingauthor]{Corresponding author}
\ead{roberto_zenit@brown.edu}

\address[mymainaddress]{School of Engineering, Brown University, 184 Hope St, Providence, RI 02912, USA}
\address[otheraddress]{Instituto de Investigaciones en Materiales,Universidad Nacional Autónoma de México, México}

\begin{abstract}
We experimentally investigate the interaction between a pair of bubbles ascending in a stagnant viscoelastic shear-thinning fluid. In particular, we focus on the effect of bubble size, across the velocity discontinuity, on the bubble-bubble interaction. Compared to the drafting-kissing-tumbling (DKT) behavior in Newtonian fluid, bubbles in the viscoelastic shear-thinning fluid exhibit, what we call, drafting-kissing-dancing (DKD) phenomenon. In the dancing phase, the bubble pair repeatedly interchange their relative leading and trailing positions as they rise to the free surface. To gain further insights, the flow fields around the bubble pair interaction are obtained using particle image velocimetry (PIV). From the experimental results, we suggest that the elasticity, deformability, and negative wake are responsible for such an interaction between the bubble pair, thus revealing the fundamental physics of bubble clustering often observed in non-Newtonian fluids.

\end{abstract}


\end{frontmatter}
%Introduction

\section{Introduction}
Motion of gas bubbles in non-Newtonian fluids has been studied since the late 1960s due to their widespread applications in the chemical processes, such as fluidized beds, bioreactors and heat transfer applications \cite{cerri2008average,kulkarni2005bubble}. The pioneering work of Astarita and Apuzzo \cite{astarita1965motion} showed that a single bubble rising in the viscoelastic fluid has a discontinuity in the bubble velocity as the bubble volume gradually increases. They suggested that the transition of the interface characteristics from Stokes (rigid) to Hadamard regime (free) and the viscoelasticity of the fluid is responsible for such an abruptness in the volume versus velocity curve. Later, Leal et al. \cite{leal1971motion} provided strong evidence to support the hypothesis of Astarita and Apuzzo by comparing the terminal velocities of approximately equal sized glass spheres and air bubbles. However, the velocity discontinuity was not observed for the glass spheres. Hassager \cite{hassager1979negative} investigated the flow fields around a rising bubble in a solution of polyacrylamide in glycerol using a laser-Doppler anemometer. He reported that behind the bubble, the flow moves in the opposite direction to that of the bubble motion, which he referred to as ``negative wake". Since these original works, the motion of a single bubble in viscoelastic fluids has been extensively studied, both experimentally and numerically.

Herrera-Velarde et al. \cite{herrera2003flow} experimentally found drastic changes in the flow field downstream of the rising bubble below and above the critical volume. Using particle image velocimetry, they showed that a negative wake is formed only for the supercritical bubbles. Further, their study concluded that the magnitude of the bubble velocity is susceptible to the wall effects. However, the critical bubble volume at which the velocity discontinuity occurs remains undisturbed. In their numerical study, Pillipakkam et al. \cite{pillapakkam2007transient} using level set method and Oldroyd-B as the constitutive equation predicted the velocity discontinuity above the critical bubble volume. They suggested that a negative wake (which acts as an upward thrust to the bubble) is responsible for the increase in the terminal velocity of the bubble. However, for spheres sedimenting in viscoelastic fluids, negative wakes were observed without an abrupt discontinuity in the velocity \cite{arigo1998experimental}. Moreover, Velez-Cordero et al. \cite{velez2012study} showed that bubbles in Boger fluids (fluids which have elasticity but with a constant shear viscosity) exhibit the velocity discontinuity without the presence of a negative wake. More recently, Fraggedakis et al. \cite{fraggedakis2016velocity} made a notable contribution in understanding the bubble velocity discontinuity in viscoelastic fluids. Using pseudo arc-length continuation in their numerical study, they showed that there exist two hysteresis loops, which serves as a continuation to the separate branches of the discontinuity. The first (small) hysteresis loop corresponds to the deformation of the bubble shape and the appearance of a negative wake. Whereas the second hysteresis loop, assisted by the change in the rheological properties in front of the bubble, is responsible for the abrupt increase in the bubble's velocity. In their results, they observed that for strong shear-thinning fluids the two hysteresis loops merge. Thus, the bubble shape deformation, negative wake formation, and the velocity jump occur simultaneously. These results agreed with the experimental results by Herrera-Velarde et al. \cite{herrera2003flow}, Velez-Cordero et al. \cite{velez2012study}, and Pilz and Brenn \cite{pilz2007critical}. 

Recently, Bothe et al. \cite{bothe2022molecular} identified the molecular mechanism behind the formation of velocity discontinuity across the subcritical and supercritical bubbles. From their investigations, they showed that the rise velocity jump discontinuity is the result of interplay between the convection and relaxation and their characteristic time scales. As the bubble rises, the polymer molecules are oriented and stretched in the circumferential direction near the upper pole of the bubble. This molecular orientation can either relax before the equator (subcritical case) or beyond the equator (supercritical case). All these investigations give a complete understanding of single bubble motion in viscoelastic shear-thinning fluids. Nevertheless, only very little has been explored in the bubbly flows of non-Newtonian fluids, especially viscoelastic shear-thinning fluids \cite{zenit2018hydrodynamic}. 

Velez-Cordero et al. \cite{velez2011hydrodynamic} studied the interaction between two bubbles rising in shear-thinning inelastic fluids. Contradictory to the two-bubble interaction in Newtonian fluids, they showed that no tumbling phase occurs in non-Newtonian fluids. Further, the bubbles stayed together after the interaction, forming a doublet (or clusters). Velez-Cordero et al. \cite{velez2011bubble} provided experimental evidence to the bubble clustering in shear-thinning inelastic fluids and put forth a set of conditions for the transition from non-coalescing to coalescing flows. A similar study on the properties of the bubbly flows in the Boger-type fluids was also carried out \cite{velez2012study, velez2014compact}. 

Direct numerical simulations (DNS) showed that two bubbles rising inline in a viscoelastic fluid can either form a stable chain or coalesce with each other \cite{yuan2021hydrodynamic}. Extending this study to chain of bubbles, Yuan et al. \cite{yuan2021vertical} suggested that the viscoelastic normal stresses are responsible for the clustering of bubbles in non-Newtonian fluids. Studying these bubble clusters is essential, as they enhance bubble coalescence, which induces a premature transition to the churn-turbulent flow in non-Newtonian fluids. Although there has been a little focus on the numerical studies of non-Newtonian bubbly flows, all the experimental studies so far are compromised by neglecting either the elastic effects or the shear-thinning effects to understand the bubble-bubble interaction.

In the present study, the hydrodynamic interaction between an identical bubble pair rising inline in the viscoelastic shear-thinning fluid is studied experimentally. We report that across the critical bubble volume, the bubble-bubble interaction significantly changes. Subcritical bubble pair showed drafting-kissing process before coalescing. Whereas, the supercritical bubble pair repeatedly switch their relative leading and trailing positions (referred here to as dancing). Flow field measurements around the bubble pair suggests that this unique drafting-kissing-dancing phenomenon is because of the negative wake, deformability of the bubbles and elasticity of the fluid. We also show that, by altering the rheological properties of the fluid, the subcritical bubble pair can have a similar drafting-kissing-dancing behavior. For contrast, the results are compared with that of the Newtonian and shear-thinning inelastic fluids.

\section{Experimental setup}

\subsection{Column and bubble generation}

\begin{figure}[H]
    \centering
    \includegraphics[scale=0.5]{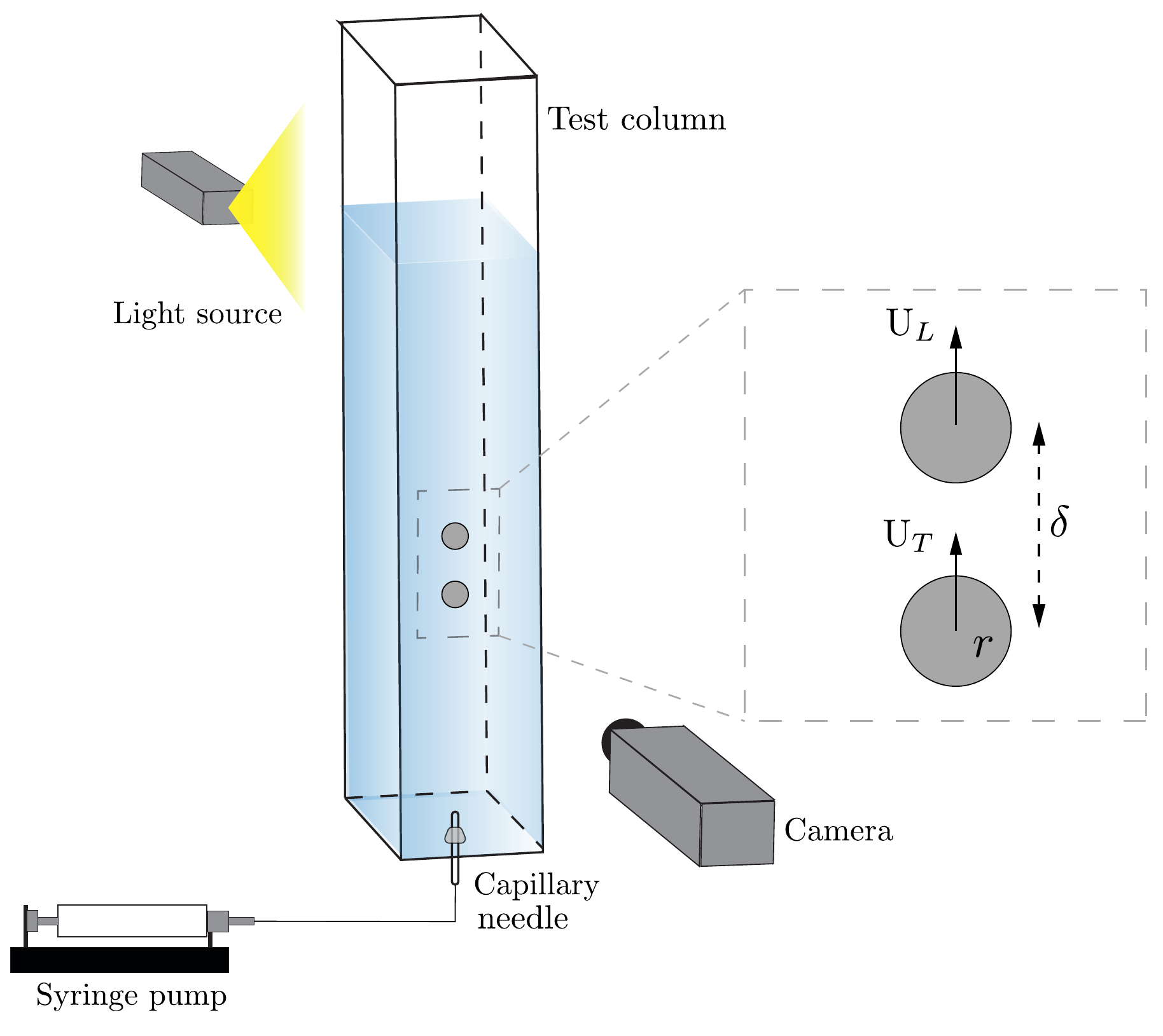}
    \caption{Experimental setup}
    \label{fig:Experimental Setup}
\end{figure}

The experiments to analyze the interaction of a bubble pair in viscoelastic shear-thinning fluids were carried out using the setup as shown in Fig. \ref{fig:Experimental Setup}. The test column made of transparent acrylic sheets with dimensions 50 mm $\times$ 50 mm $\times$ 400 mm,  adequately wide to neglect the wall effects and sufficiently high for the bubbles to reach their terminal velocities, was used. A syringe pump (\textit{Harvard Apparatus PHD 2000 series}) with a capillary needle inserted at the bottom of the test column through a self-healing rubber stopper was used to make bubbles. To generate identical inline bubble pair and to avoid the burst of gas bubbles, longer capillary needles (25.4 mm) were employed. The initial separation distance between the two bubble centers, $\delta$, was controlled using the flow rate of the syringe pump. The repeatability of the bubbles produced was ensured from the volume versus velocity curve of a single bubble. A time lapse of approximately 5 minutes was left in between the experiments to ensure that there was no residual impact of the past experiments on the current one.

\subsection{Test fluids}

To prepare viscoelastic shear-thinning fluids, Separan AP30, an anionic polyacrylamide polymer (\textit{PAAm, Sigma Aldrich with molecular weight, $M_w = 5 \times 10^6$} g mol$^{-1}$) was used. Three different concentrations of PAAm (0.15\%, 0.20\%, and 0.25\% by weight) in water were used to make test fluids with different rheological properties. These PAAm concentrations are widely studied and reported to exhibit both the shear-thinning and the viscoelastic properties \cite{leal1971motion,herrera2003flow,zana1978dynamics}. The polymers were mixed in the water using an over-head mixer at room temperature until the polymers were completely dissolved. After mixing, the fluids were left undisturbed for at least 48 hours so that the solution is homogeneous. A mixture of 90\% glycerin and 10\% water by weight was used as the benchmark Newtonian fluid. The physical properties of the fluids used are summarized in Table \ref{table:FluidProperties}. To avoid the coalescence in some experiments, Magnesium Sulfate (\textit{MgSO$_4$, Eisen-Golden Laboratories}) salt was added to one of the fluids. MgSO$_4$ was chosen because of its ability to delay the coalescence while present only in small amounts \cite{lessard1971bubble}.

\begin{table}[h]
\centering
\begin{tabular}{ p{1.25cm} p{4.35cm} p{1.85cm} p{1.35cm} p{1.35cm} p{2.25cm} p{1.35cm} p{1.35cm}}
\hline
Fluids & Composition & Symbol & $\rho$ & $\sigma$ & $\mu$ & $n$ & $\lambda$\\
& (by weight) & & (kg/m$^{3}$) & (mN/m) & (Pa s) & & (s)\\
\hline
N & 90\% Glycerin - 10\% Water & ( --- , - - ) &1222.6 & 71.89 & 0.11 & 1.0 & -\\
VE1 & 0.15\% PAAm in Water & ( \tikzcircle[fill=blue]{4pt} , \tikzcircle[draw=blue]{4pt} ) & 997.0 & 77.08 & 2.26 - 0.014 & 0.21 & 5 \\
VE2 & 0.20\% PAAm in Water & ( \tikzsquare[fill=red]{8pt}, \tikzsquare[draw=red]{8pt}) &997.4 & 77.83 & 4.89 - 0.017 & 0.10  & 12 \\
VE3 & 0.25\% PAAm in Water & ( \tikztriangle[fill=green]{8pt}, \tikztriangle[draw=green]{8pt}) &998.8 & 78.31 & 6.70 - 0.024 & 0.08 & 39 \\
VE1-S & 0.15\% PAAm in Water + 0.0625\% MgSO$_4$ & ( \tikzdiamond[fill=magenta]{10pt}, \tikzdiamond[draw=magenta]{10pt}) & 997.8 & 76.87 & 0.98 - 0.013 & 0.34 & 2 \\
\hline
\end{tabular}
\caption{Physical properties of the fluids: \\ $\rho$ - density; $\sigma$ - surface tension; $\mu$ - viscosity; $n$ - power index; $\lambda$ - relaxation time }
\label{table:FluidProperties}
\end{table}

\begin{figure}[H]
    \begin{subfigure}[h]{0.5\textwidth}
    \centering
    \includegraphics[scale=0.5]{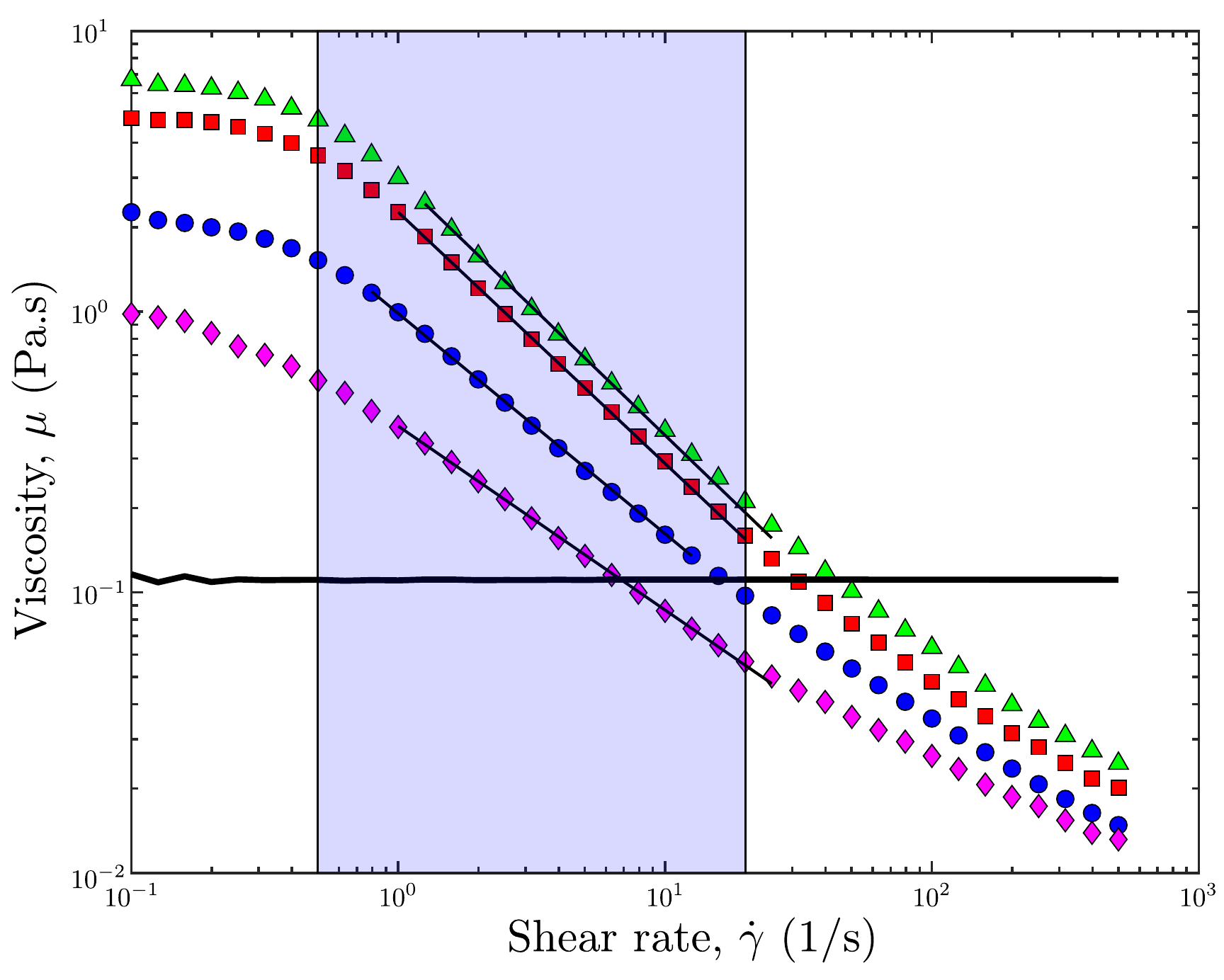}
    \caption{}
    \label{fig:Viscosity}
\end{subfigure}
\begin{subfigure}[h]{0.5\textwidth}
\centering
    \includegraphics[scale=0.5]{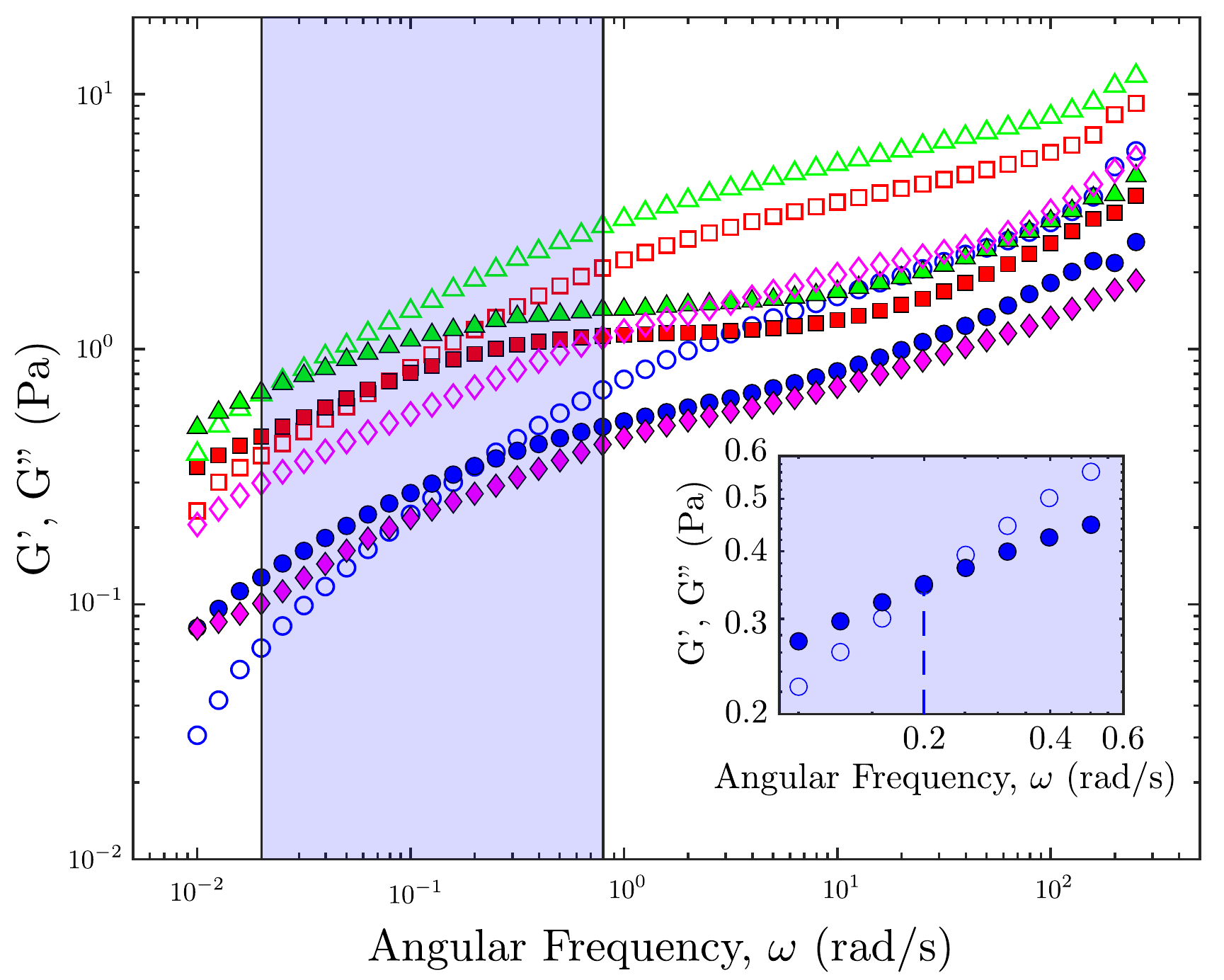}
    \caption{}
    \label{fig:Oscillation}
\end{subfigure}
    \caption{Rheology of the fluids used: (\ref{fig:Viscosity}) Shear viscosity as a function of shear rate for N ( --- ), VE1 ( \tikzcircle[fill=blue]{4pt} ), VE2 ( \tikzsquare[fill=red]{8pt}),  VE3 ( \tikztriangle[fill=green]{8pt}), and VE1-S ( \tikzdiamond[fill=magenta]{10pt}). The bold lines show the fitting of the power-law model for different percentages of PAAm;  (\ref{fig:Oscillation}) G$''$, Loss modulus and G$'$, storage modulus as a function of angular frequency.  The closed symbols ( \tikzcircle[fill=blue]{4pt} , \tikzsquare[fill=red]{8pt}, \tikztriangle[fill=green]{8pt}, \tikzdiamond[fill=magenta]{10pt}) corresponds to the loss modulus and the open symbols  ( \tikzcircle[draw=blue]{4pt} ,  \tikzsquare[draw=red]{8pt},  \tikztriangle[draw=green]{8pt}, 
     \tikzdiamond[draw=magenta]{10pt}) corresponds to the storage modulus of VE1, VE2, VE3, and VE1-S fluids, respectively. Fig. (\ref{fig:Oscillation}) inset shows the angular frequency at which the loss modulus ( \tikzcircle[fill=blue]{4pt} ) and storage modulus ( \tikzcircle[draw=blue]{4pt} ) intersect for VE1 fluid.}
    \label{fig:ViscosityandOscillation}
\end{figure}

The rheological properties of the fluids were measured with a rheometer (\textit{ARES-G2 Rheometer, TA instruments}) using a cone-plate geometry (40 mm, 0.04 rad cone plate angle with a gap of 24 $\mu$m). The surface tension of the fluids was measured with a bubble pressure tensiometer (\textit{BPT Mobile, KRUSS Scientific Instruments}). The shear and oscillatory tests of the aqueous PAAm fluids are shown in Fig. \ref{fig:Viscosity} and Fig. \ref{fig:Oscillation}, respectively. The blue shaded region in Fig. \ref{fig:Viscosity} and Fig. \ref{fig:Oscillation} corresponds to the range of shear rates in the bubble rising experiments (0.5 to 20 $1/s$) in VE1. This shows that the experiments were carried out in the shear-thinning regime and in both the viscous dominant and elastic dominant regime of VE1. For most fluids the relaxation time, $\lambda$, was obtained from the angular frequency, $\omega$, at which G' and G'' intersect.

Since there was no intersection of the storage modulus and loss modulus for VE1-S, as shown in Fig. \ref{fig:Oscillation}, the generalized Maxwell model was used to fit the experimental values of the G$'(\omega$) and G$''(\omega$) as reported in the literature \cite{liu2011force,espinosa2013fluid}.
The storage modulus and loss modulus are given by,
\begin{equation}
   G'(\omega) = \sum_{i=1}^{N} \frac{g_i\lambda_i^2\omega^2}{1+\lambda_i^2\omega^2}, \hspace{1.25cm}  \text{and} \hspace{1.25cm} G''(\omega) = \omega\mu + \sum_{i=1}^{N} \frac{g_i\lambda_i\omega}{1+\lambda_i^2\omega^2}.
    \label{Eq:G'G''}
\end{equation} The values of the fitting parameters ($g_i$ and $\lambda_i$) are used to calculate the mean relaxation time, 
\begin{equation}
   \lambda = \sum_{i=1}^{N} \frac{g_i\lambda_i^2}{g_i\lambda_i}.
    \label{Eq:RelaxationTime}
\end{equation} For N = 4, a good fit for the G$'$ and G$''$ is achieved, leading to a $\lambda$ = 2 s for this fluid.  

\subsection{Bubble size and velocity measurement}

The bubble radius and the bubble velocity were measured using a high-speed camera (\textit{Photron FASTCAM SA5}) with a 500 frames/s recording rate. The camera was positioned such that the field of view covers the entire bubble column. A 4 mm $\times$ 4 mm checked square board was used as a calibration image. To capture the slightest changes in the bubble's shape, camera with a magnification lens was used to record the bubble pair interaction at 3000 frames/s. An LED light source, placed directly behind the bubble column facing the camera direction, was used as a diffused back light. 

The recorded images were analyzed using MATLAB. To find the bubble radius, the background from the raw image was subtracted and converted to a binary image using an appropriate threshold value. From the binary image, the equivalent spherical bubble radius, $r$, was calculated as,

\begin{equation}
    r = \frac{({d_{\text{max}}^{2}}{d_{\text{min}}^{ }})^{1/3}}{2} ,
    \label{Eq:BubbleRadius}
\end{equation} where $d_{\text{max}}$ is the longest bubble diameter and $d_{\text{min}}$ is the shortest bubble diameter.

To find the bubble velocity, the position of the bubble was determined from the binary image and its respective centroid was calculated. Then, the centroid of the same bubble was identified in the subsequent image. Since the time between each subsequent frame is known, the displacement of the centroid gives the bubble velocity. This process is repeated for the entire image sequence to find the terminal velocity of the bubble using a central difference scheme. In this manner, the uncertainty in the measured velocity was reduced. For the case of bubble-bubble interaction, the same procedure was employed to individually determine the velocities of each bubble, respectively. 

\subsection{Flow fields around the bubbles}

The flow fields around the bubbles were obtained using particle image velocimetry (PIV). Laser beam (532 nm) from a Nd:YAG laser system, generated into a thin sheet using a cylindrical lens, was used for illuminating the bubble column from right to left. To avoid the Mie-scattering from the bubble interface, the 55 $\mu$m diameter tracer particles were dyed with Rhodamine B \cite{herrera2003flow,bothe2022molecular}. The tracer particles in Ethanol solution of Rhodamine B was heated slightly above the room temperature for 24 hrs to allow for the dye to infiltrate into the tracer particles. This ensures that the dye is dispersed homogeneously along the tracer particle's surface. The dyed tracer particles were then washed thoroughly in water before mixed with the VE1 fluid. When illuminated by a green laser sheet, the dyed tracer particles emit orange color fluorescence. Using an orange filter in the camera, only the fluorescence light from the tracer particles was recorded to obtain the flow fields around the bubbles. For PIV measurements of bubble pair, great care was taken to make sure the bubbles lie in the mid of the laser sheet. The recorded images were then analyzed using PIVLab in MATLAB \cite{thielicke2021particle}.  

\section{Results and discussion}

\subsection{Single bubble velocity discontinuity}

\begin{figure}[H]
    \begin{subfigure}[h]{0.51\textwidth}
    \centering
    \includegraphics[scale=0.5]{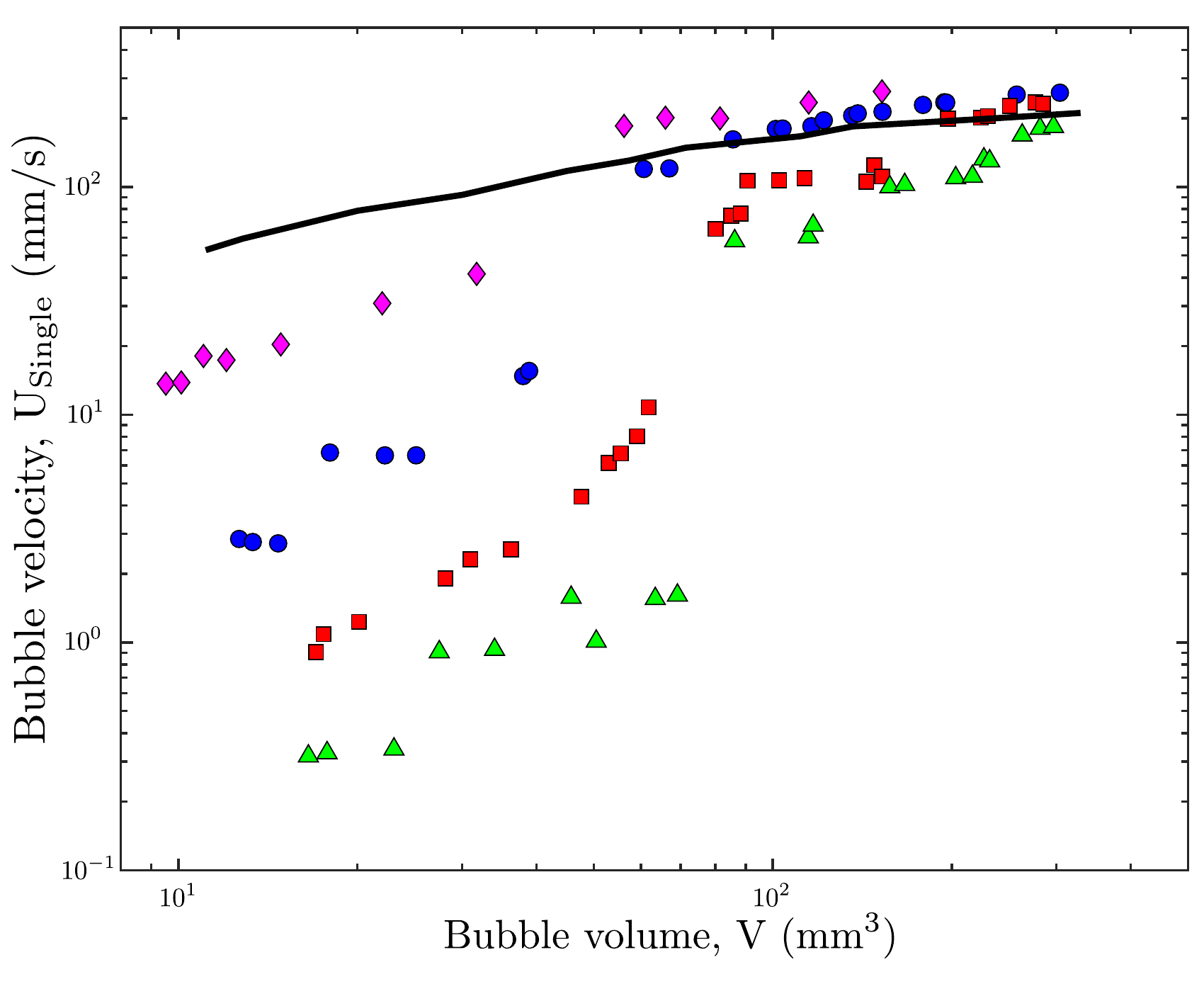}
    \caption{}
    \label{fig:VelocityDiscontinuity}
\end{subfigure}
\begin{subfigure}[h]{0.51\textwidth}
\centering
    \includegraphics[scale=0.5]{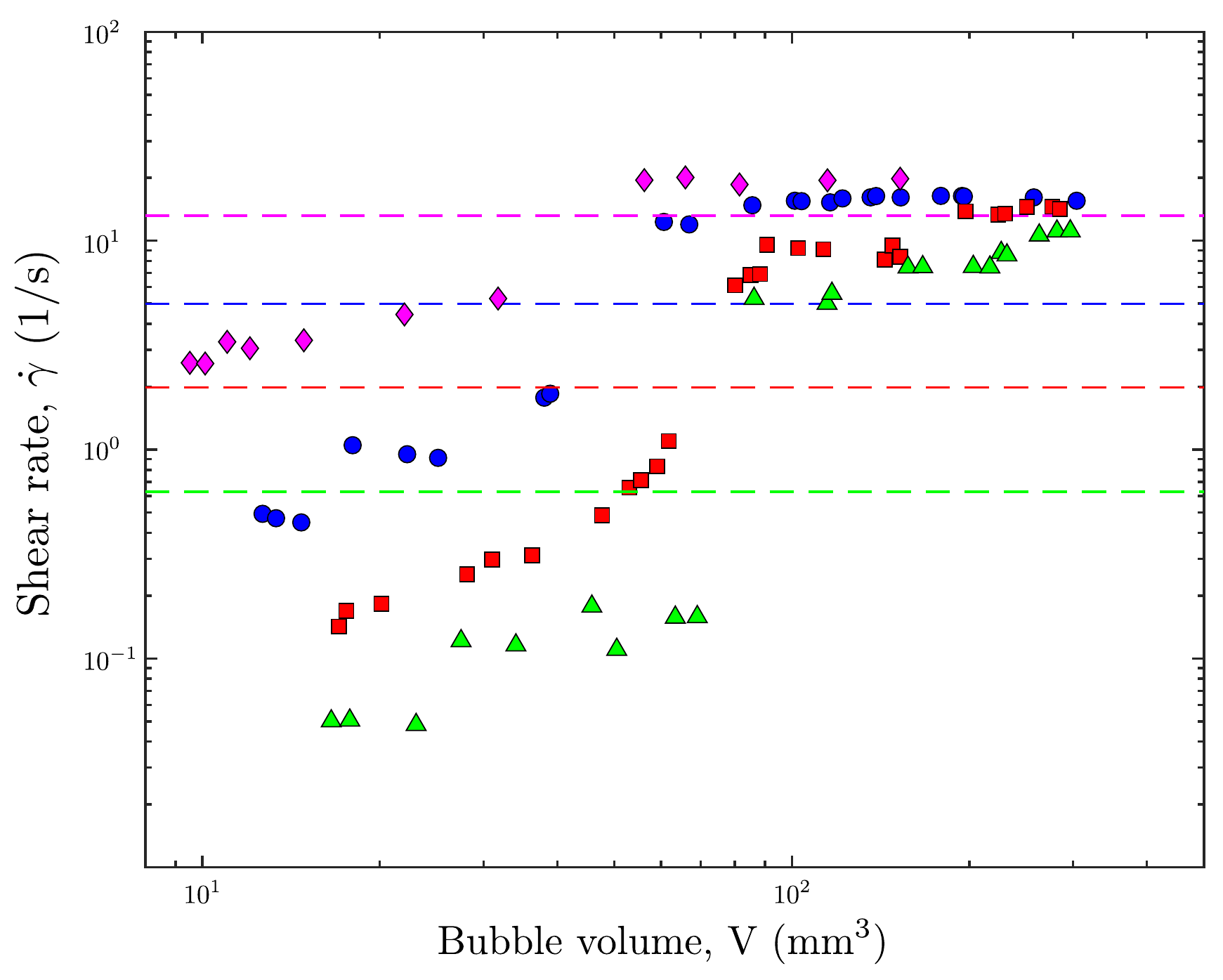}
    \caption{}
    \label{fig:ShearDiscontinuity}
\end{subfigure}
 \caption{(\ref{fig:VelocityDiscontinuity}) Rising velocity and;  (\ref{fig:ShearDiscontinuity}) Rising shear rate of a single bubble as a function of bubble volume in N ( --- ), VE1 ( \tikzcircle[fill=blue]{4pt} ), VE2 ( \tikzsquare[fill=red]{8pt}), VE3 ( \tikztriangle[fill=green]{8pt}), and VE1-S ( \tikzdiamond[fill=magenta]{10pt}). The dashed lines represent the shear rate at which the storage modulus and loss modulus intersect for different concentrations of PAAm. The error bars are of the marker size.}
\end{figure}

A clear discontinuity in the terminal bubble velocity was observed by gradually increasing the bubble volume, as shown in the Fig. \ref{fig:VelocityDiscontinuity}. With the increase in polymer concentration, the terminal bubble velocity decreases. This is attributed in part to the increased viscosity. The magnitude of velocity discontinuity, i.e. the ratio of subcritical bubble velocity to the supercritical bubble velocity, is 7.4 for VE1, 6.1 for VE2 and 4.4 for VE1-S \cite{fraggedakis2016velocity,pilz2007critical}. For VE1-S fluid, the bubble velocity is higher than that of in VE1 fluid. This is because the addition of MgSO$_4$ salt not only increased the storage modulus of the fluid in the lower shear rates (Fig. \ref{fig:Oscillation}) but also reduced the shear viscosity (Fig. \ref{fig:Viscosity}). Further, the discontinuity in shear rate of the rising bubble, $\dot\gamma$ = $U_{\text{Single}}/r$, demonstrated by Soto et al. \cite{soto2006study}, lie at the intersection of the storage and loss moduli as shown in the Fig. \ref{fig:ShearDiscontinuity}.

\begin{figure}[H]
    \begin{subfigure}[h]{0.5\textwidth}
    \centering
    \includegraphics[scale=0.35]{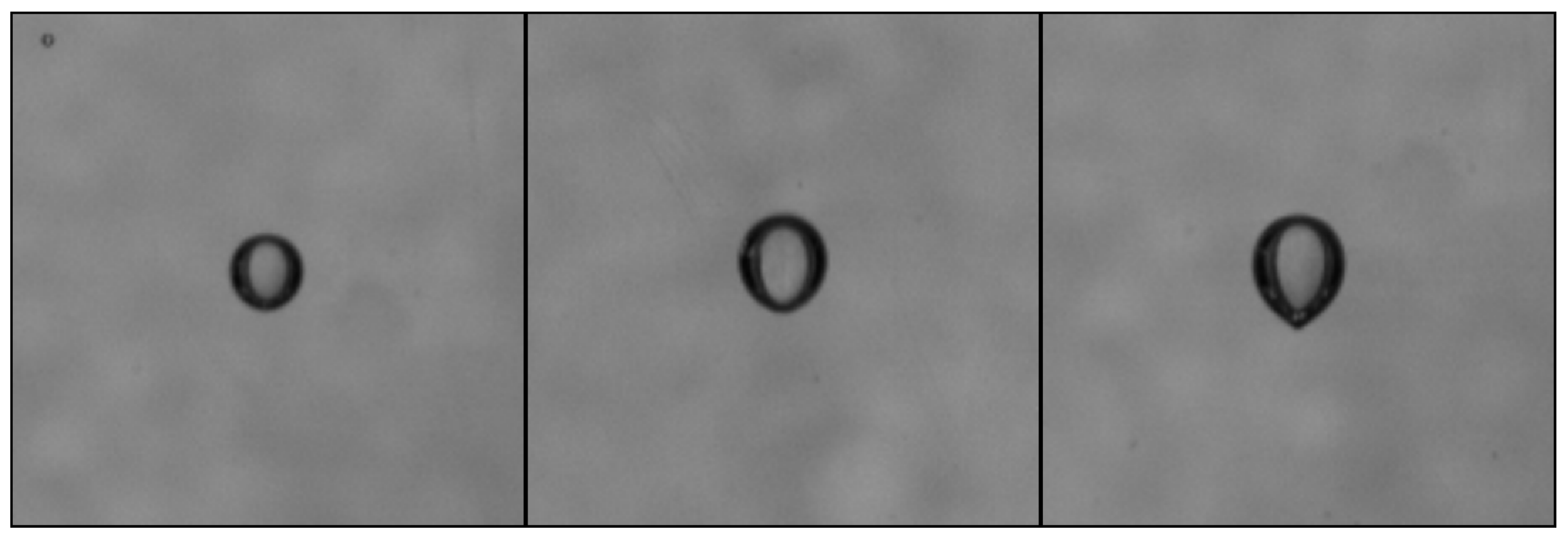}
    \caption{}
    \label{fig:BeforeCriticalVolume}
\end{subfigure}
\begin{subfigure}[h]{0.5\textwidth}
    \centering
    \includegraphics[scale=0.35]{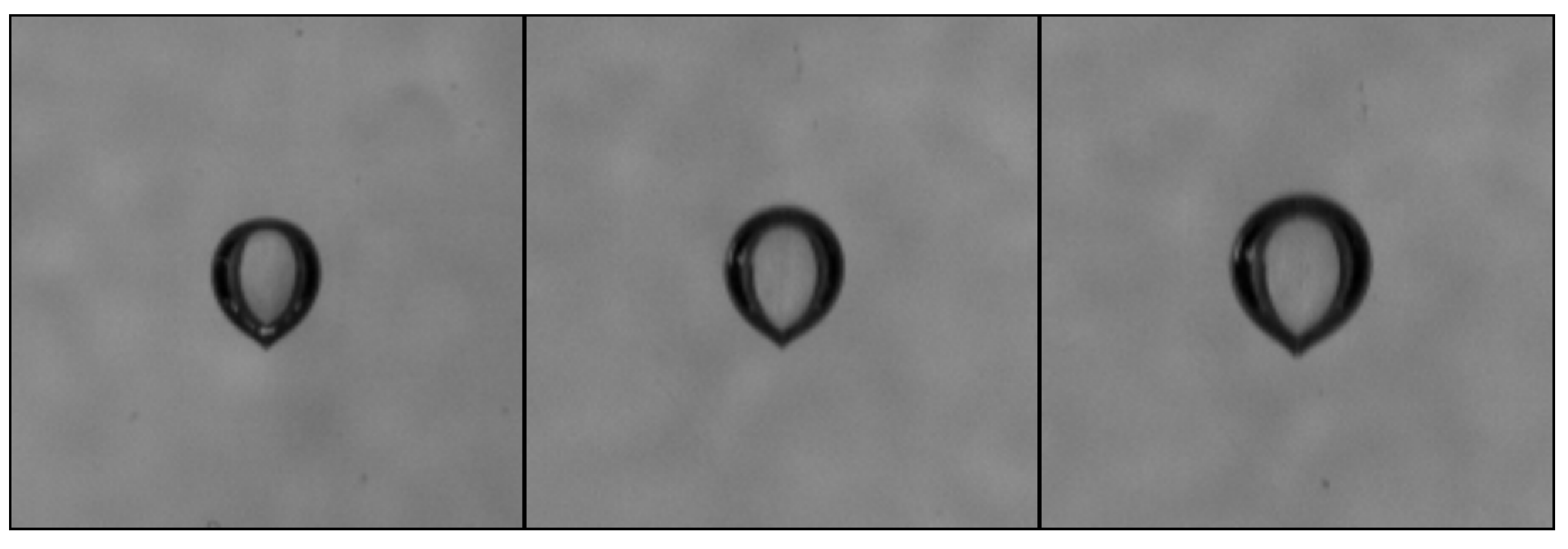}
    \caption{}
    \label{fig:AfterCriticalVolume}
\end{subfigure}
 \caption{Bubble shapes in VE2 fluid: (\ref{fig:BeforeCriticalVolume}) for subcritical bubble volumes - 12 mm$^3$, 27 mm$^3$, 45 mm$^3$  (from left to right) and;  (\ref{fig:AfterCriticalVolume}) supercritical bubble volumes - 57 mm$^3$, 68 mm$^3$ and 92 mm$^3$ (from left to right) in a window of size 18 mm $\times$ 18 mm.}
\end{figure}

\begin{figure}[H]
    \begin{subfigure}[h]{0.5\textwidth}
\centering
    \includegraphics[scale=0.25]{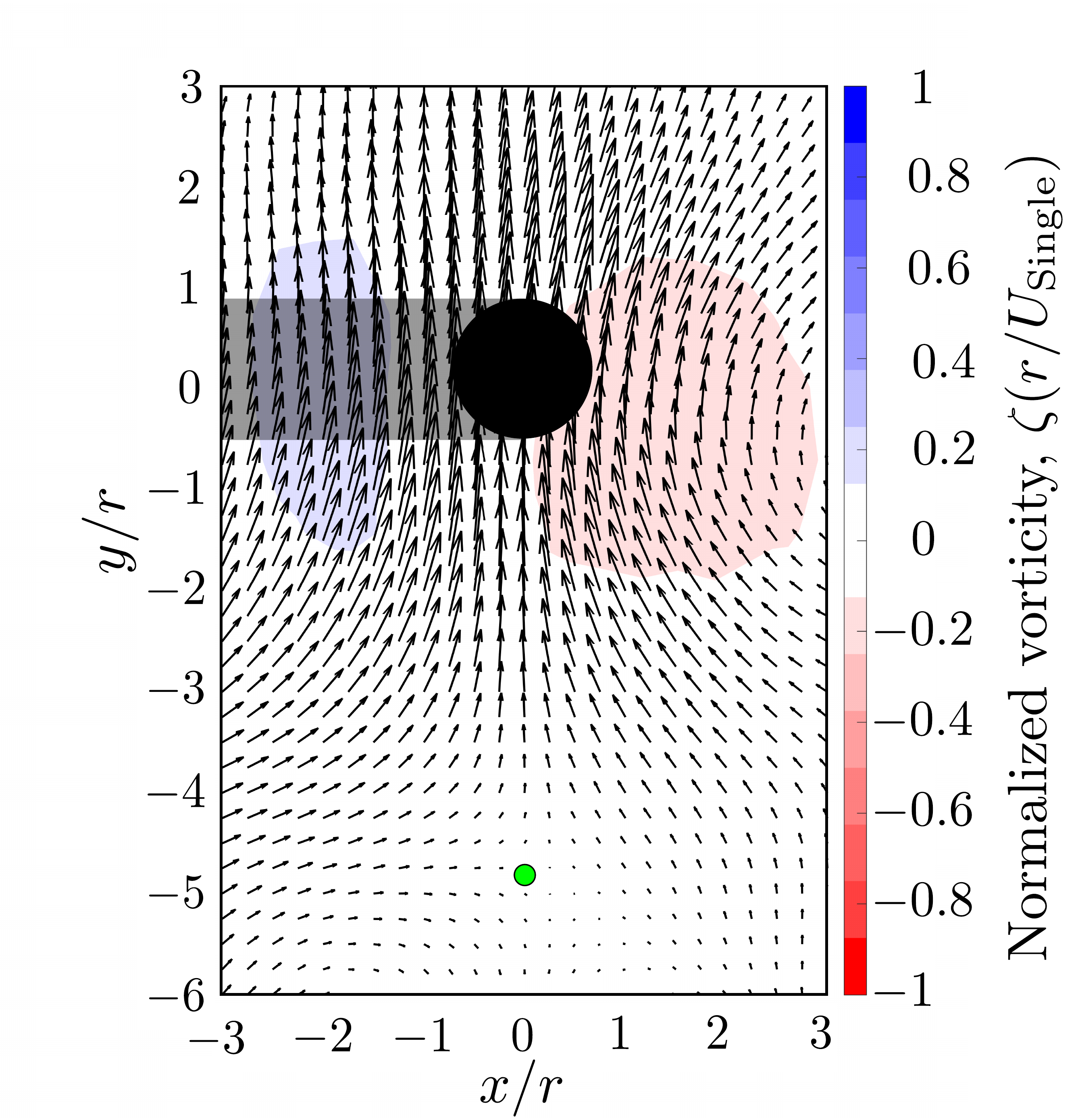}
    \caption{}
    \label{fig:PositiveWake}
\end{subfigure}
\begin{subfigure}[h]{0.51\textwidth}
\centering
\centering
    \includegraphics[scale=0.25]{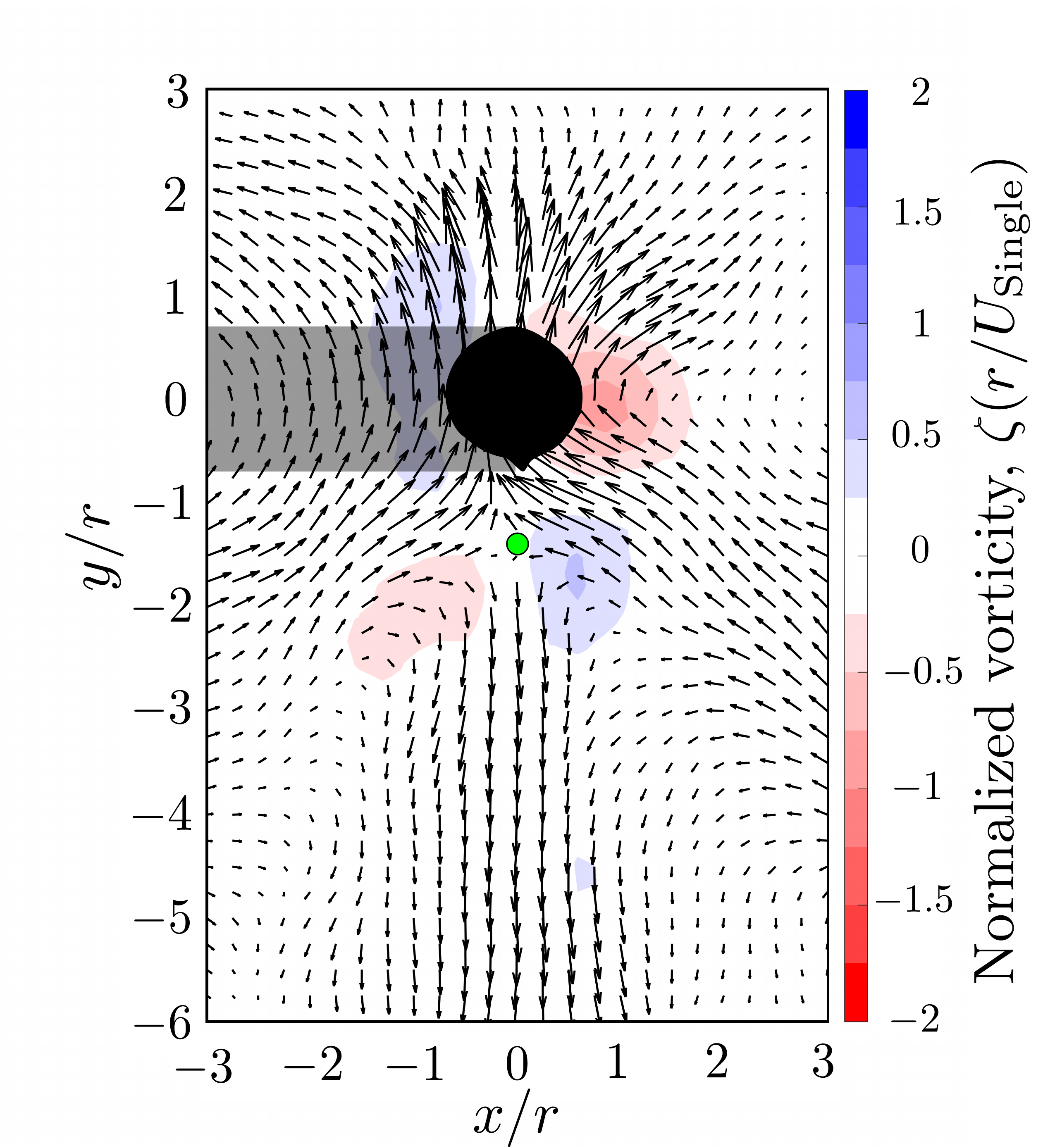}
      \caption{} 
    \label{fig:NegativeWake}
    \end{subfigure}
     \caption{Instantaneous velocity fields around a single bubble rising in the VE1 fluid: (\ref{fig:PositiveWake})  Subcritical volume. The Reynolds number and Weissenberg number are 0.25 and 8, respectively; (\ref{fig:NegativeWake}) Supercritical volume. The Reynolds number and Weissenberg number are 5 and 220, respectively. The center of the bubble is at (0, 0) and the coordinates are normalized by the bubble radius. The contours depict the normalized vorticity fields around the bubbles. The stagnation point in the wake of the bubble is indicated by a green dot.}
\end{figure}

Similar to numerous experimental and numerical studies \cite{herrera2003flow,pillapakkam2007transient,fraggedakis2016velocity,pilz2007critical,zenit2018hydrodynamic}, a sharp change in the bubble shape across the velocity discontinuity was observed as seen in Fig. \ref{fig:BeforeCriticalVolume} and Fig. \ref{fig:AfterCriticalVolume}. For subcritical bubbles, the shape is spheroidal but slightly elongated in the stream wise direction, thus adopting an inverted teardrop shape. However, the supercritical bubbles developed a pointed cusp. Fig. \ref{fig:PositiveWake} shows the velocity fields around a subcritical bubble normalized by its terminal velocity in VE1 fluid. The flow field closely resembles that of a bubble in a Newtonian fluid, nevertheless with an additional vortex ring in the wake. The contours show the vorticity, $\zeta$, normalized by the shear rate, $\dot\gamma$ = $U_{\text{Single}}/r$. The Reynolds number and the Weissenberg number are defined as, $Re$ = $2{\rho U_{\text{Single}} r}/{\mu}$, and Wi = $\lambda \dot\gamma$, respectively. The lower stagnation point, where the flow reversal occurs, lie at least 5 bubble radius away from the rear side of the subcritical bubble. Whereas, for the supercritical bubble, as seen in Fig. \ref{fig:NegativeWake}, the lower stagnation point is within one bubble radius distance away. 

\subsection{Bubble pair interaction}

To identify the effect of viscoelasticity and shear dependent viscosity on the interaction between two bubbles, we first conducted a series of experiments in two well-known scenarios (see Appendix): (1) interaction of a bubble pair in a Newtonian fluid and (2) interaction of a solid particle pair in a shear-thinning viscoelastic fluid. For the case of bubble pair interaction in a Newtonian fluid, the classical drafting-kissing-tumbling (DKT) behavior was observed for the range of Reynolds number relevant to this study \cite{kok1989dynamics,kok1993dynamics,brennen2005fundamentals}. The interaction between two solid spheres in a viscoelastic fluid with shear-thinning viscosity leads to the formation of vertically aligned pairs that remained stable after the initial interaction, in close agreement with \cite{joseph1994aggregation}. Therefore, considering the same fluids and the same range of relevant conditions, the deviations from these behaviors could therefore be attributed to either non-Newtonian properties and/or deformability of the bubbles. 

\subsubsection{Subcritical bubble pair interaction}

\begin{figure}[H]
\centering
    \includegraphics[width=9cm, height=9cm]{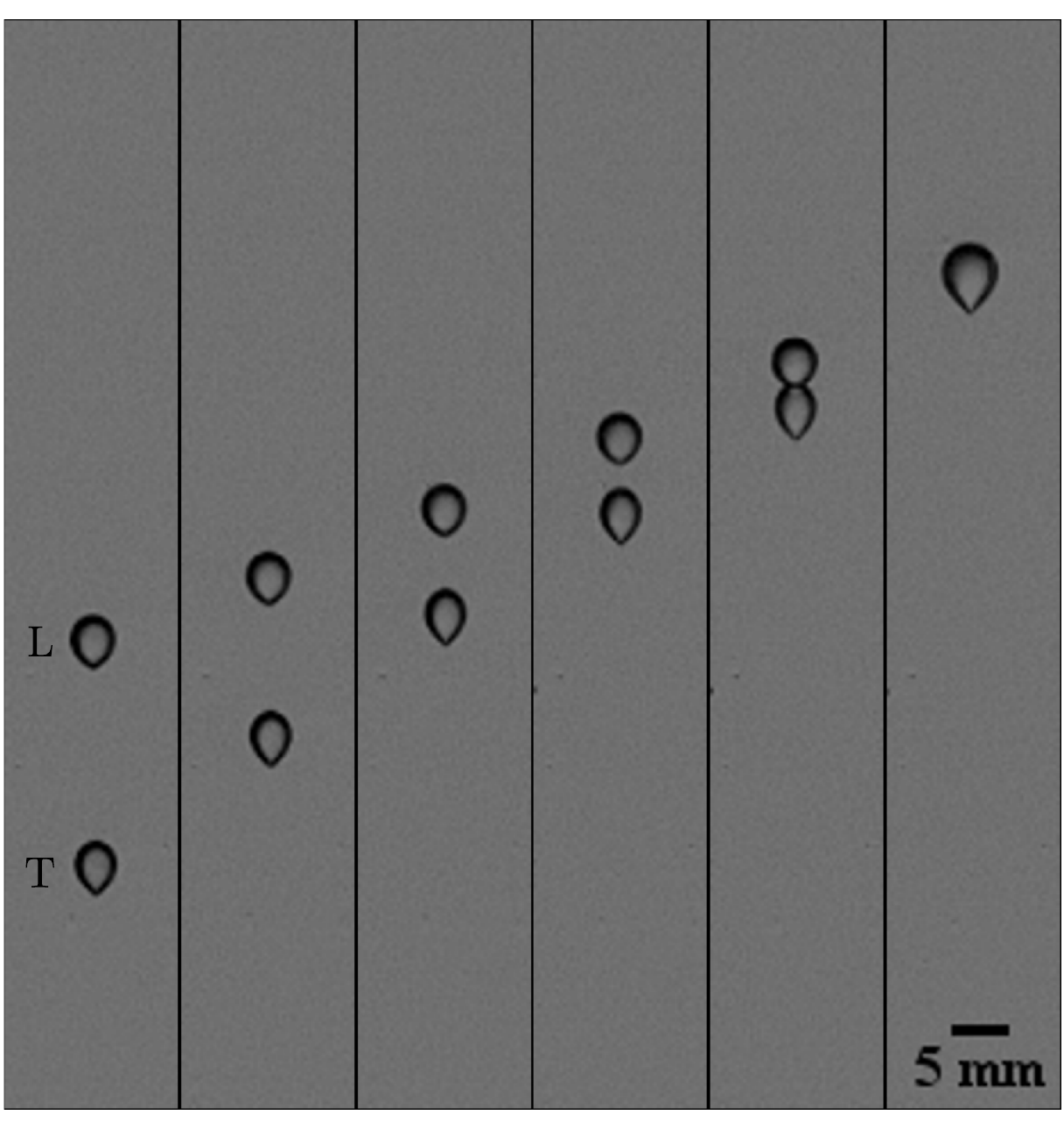}
    \caption{Series of images showing the interaction between two bubbles of volume below the critical value (with a diameter of 4 mm) in the VE1 fluid for a $Re$ $\approx$ 1 and Wi = 38. Interval between the images is 200 millisecond. L and T denote the leading and trailing bubbles at the beginning of the experiment respectively.}
    \label{fig:BCVimages}
\end{figure}

Fig. \ref{fig:BCVimages} shows the snapshots of the interaction between two subcritical bubbles in the VE1 fluid. For $Re$ $\approx$ 1, when a trailing bubble is introduced in the wake of the leading bubble, the trailing bubble accelerates and catches up with the leading bubble. This results either from the low pressure region at the wake or by the reduced viscosity corridor that appears behind the path of the leading bubble. During this drafting process, the leading bubble seems to be stable. However, the trailing bubble further deforms in the stream wise direction \cite{yuan2021hydrodynamic} and the kissing phase of the bubbles takes place. 

The interaction between the subcritical bubbles in viscoelastic shear-thinning fluids is, to this point, similar to the drafting-kissing phases observed for the two-bubble system in the Newtonian fluid and particle-particle interaction in viscoelastic fluid (see Appendix). Instead of the final tumbling phase as in Newtonian fluids, the bubbles coalesce with each other. Though the coalescence of non-identical bubbles was previously reported by De Kee et al. \cite{de1990motion}, it is important to note that they did not observe a velocity discontinuity in their study. This coalescence of subcritical bubbles in the present study contradicts with the stable equilibrium distance observed from the numerical results of Yuan et al. \cite{yuan2021hydrodynamic}. This might be the result of higher relaxation time of the fluids in the current study compared to that of the values considered in their simulations. 

\begin{figure}[H]
\centering
\subfloat[t* = -0.6]{\label{BCV_a}\includegraphics[scale=0.25]{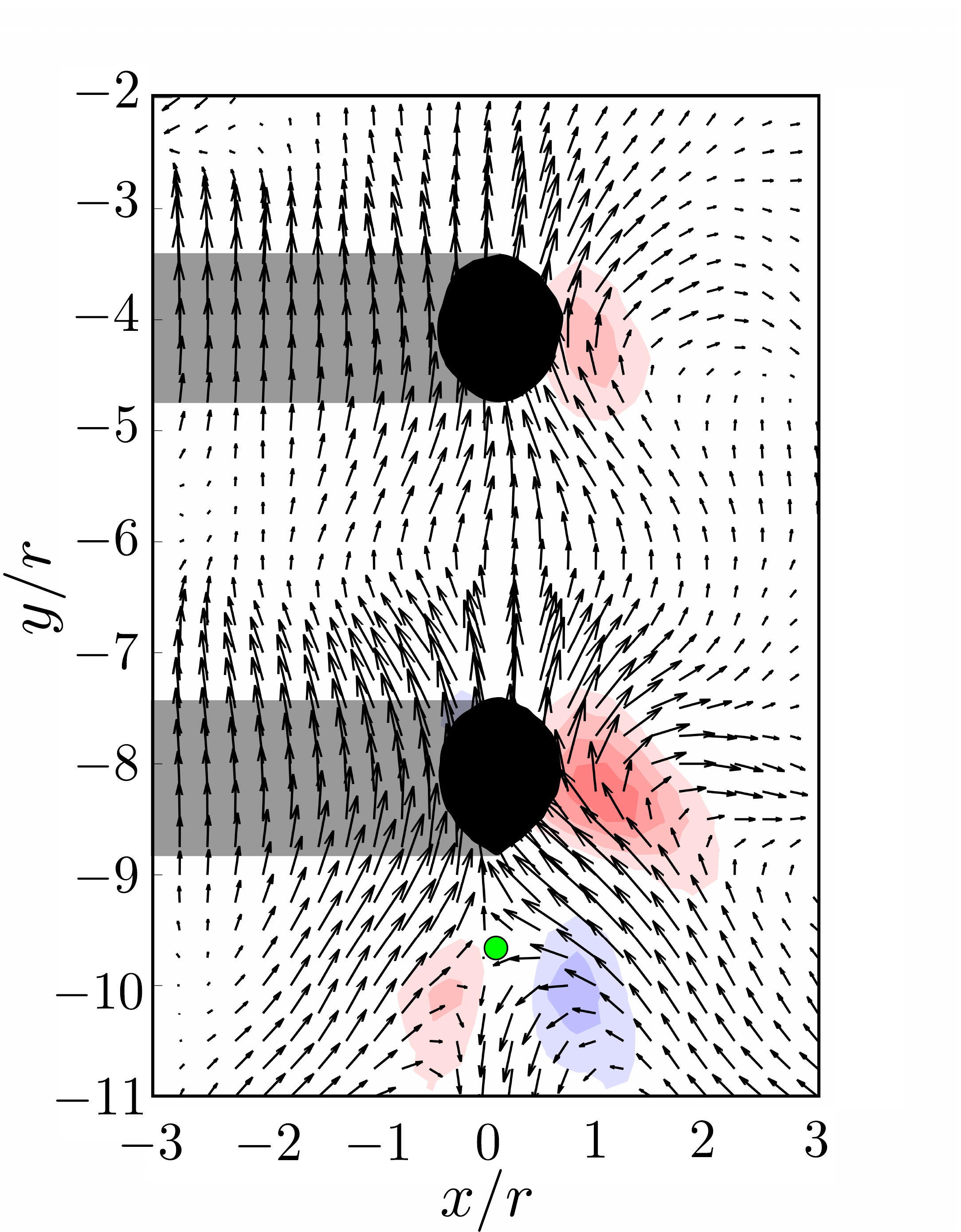}}
\subfloat[t* = -0.3]{\label{BCV_b}\includegraphics[scale=0.25]{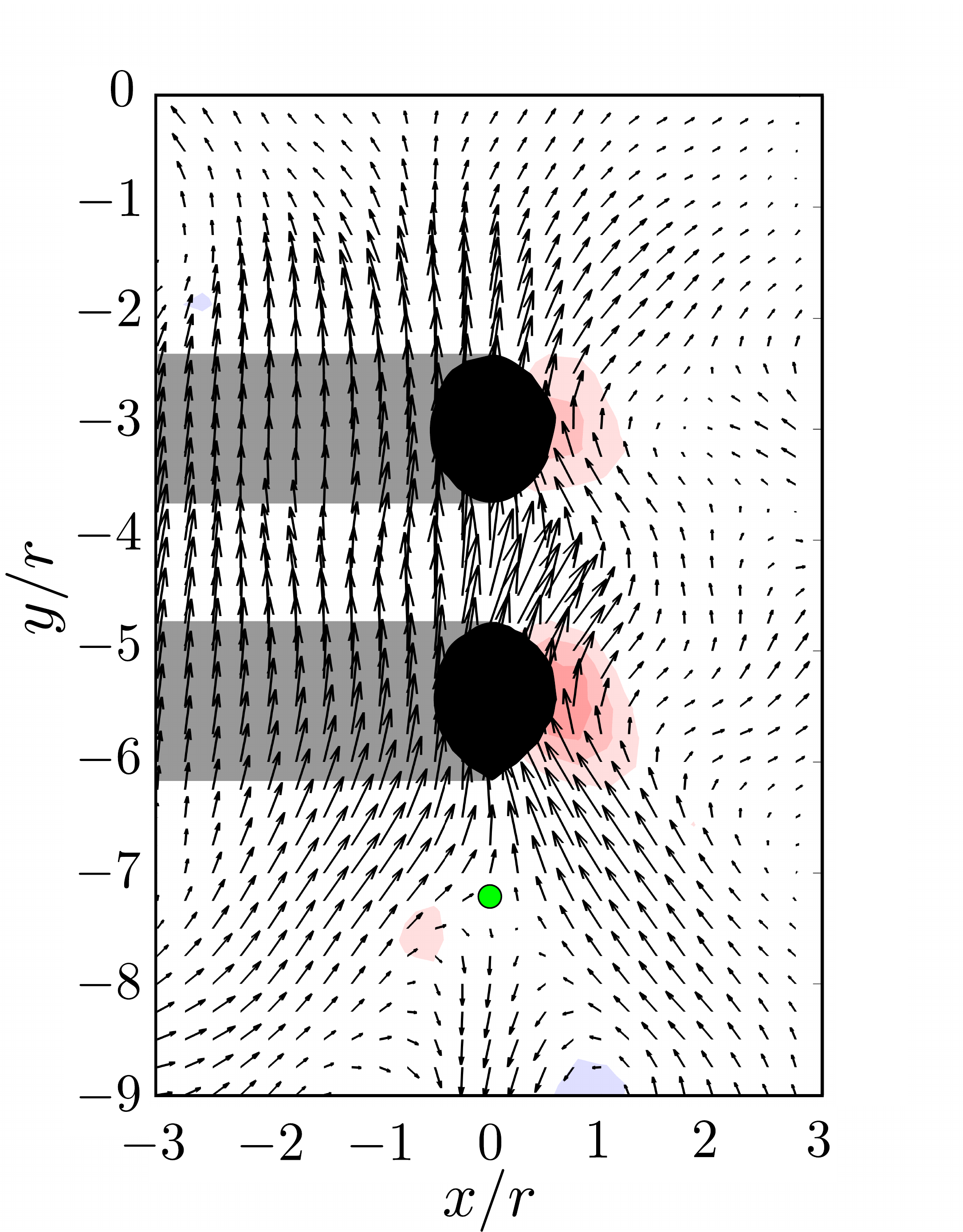}}\hfill
\subfloat[t* = -0.075]{\label{BCV_c}\includegraphics[scale=0.25]{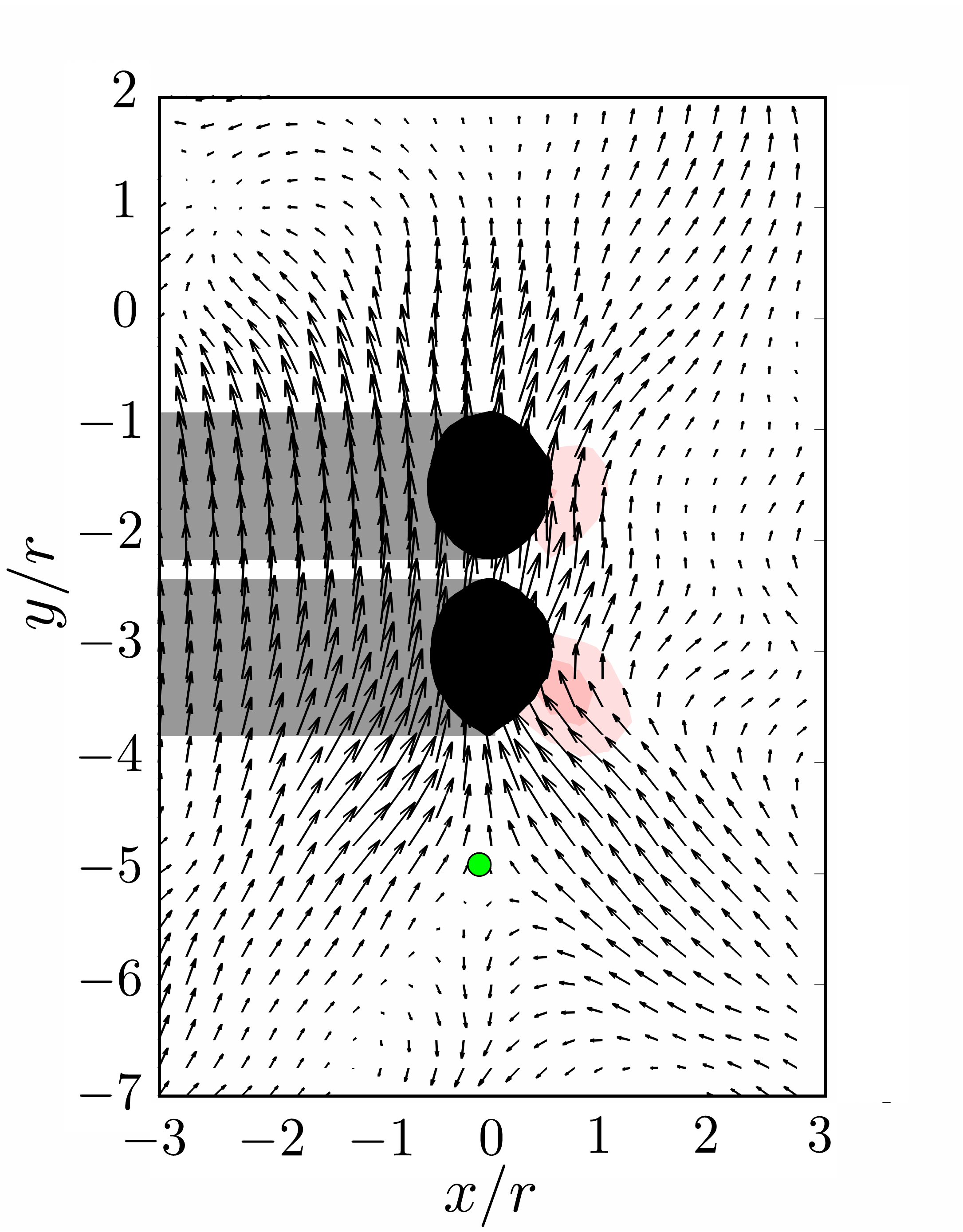}}
\subfloat[t* = 0]{\label{BCV_d}\includegraphics[scale=0.25]{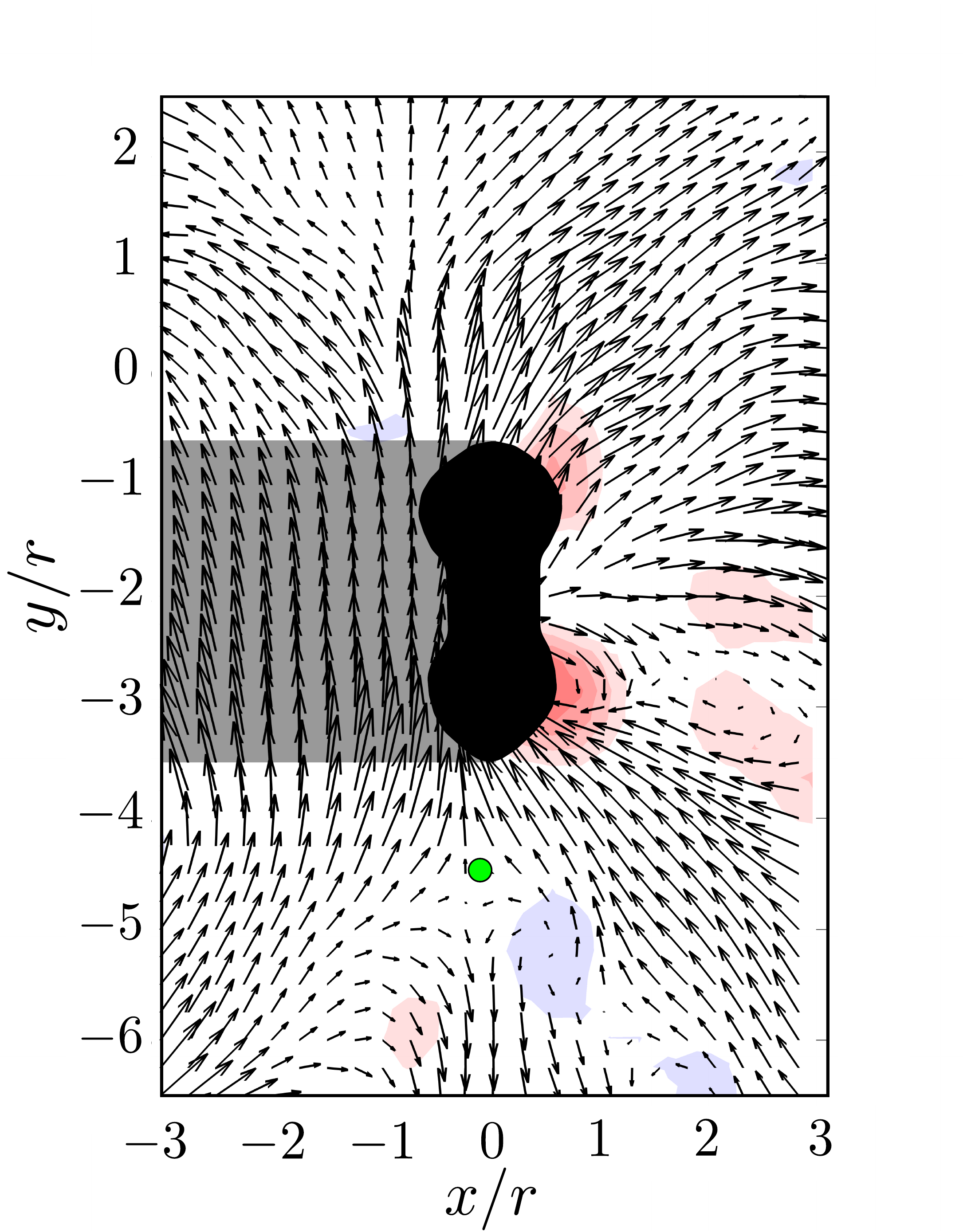}}
\subfloat[t* = 0.075]{\label{BCV_e}\includegraphics[scale=0.25]{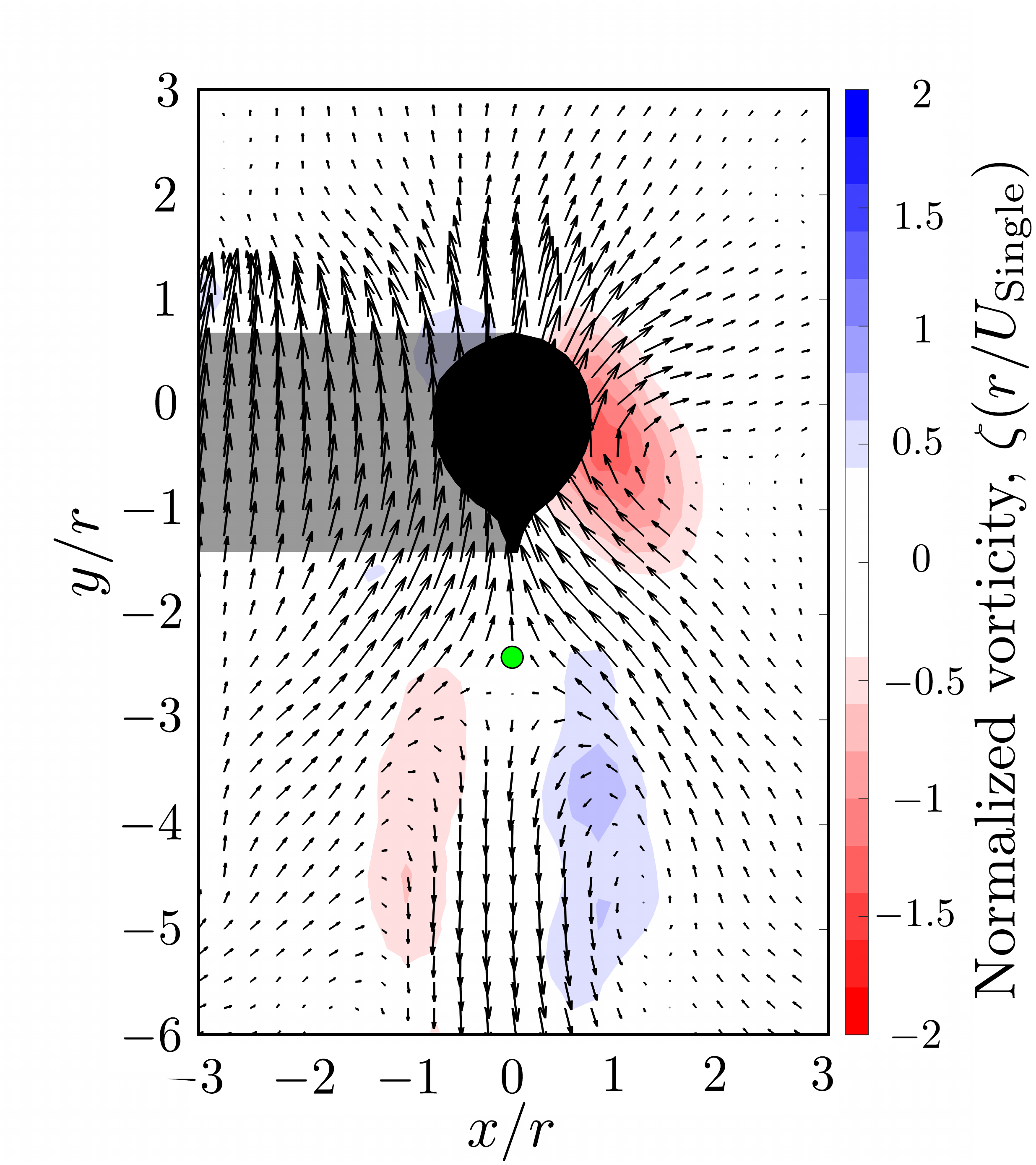}}\hfill \par
\caption{Instantaneous velocity fields around the subcritical bubble pair rising inline in the VE1 fluid. The Reynolds number and Weissenberg number are 0.35 and 25, respectively. The center of the coalesced bubble is at (0, 0) and the coordinates are normalized by the bubble radius. The contours depict the normalized vorticity fields around the bubble. The stagnation point in the wake of the bubble is indicated by a green dot.}
\label{fig:BCV_PIV}
\end{figure}

Fig. \ref{fig:BCV_PIV} depicts the instantaneous flow fields around the subcritical bubble pair interaction in the VE1 fluid over the dimensionless time, defined as $t^{*} = t(U_{\text{Single}}/r)$. The contours show the vorticity, $\zeta$, normalized by the shear rate, $\dot\gamma$ = $U_{\text{Single}}/r$. The presence of trailing bubble in the leading bubble's wake negates the weak negative wake behind the leading bubble. Though the flow field is modified, the leading bubble does not feel the presence of another bubble until the bubbles kiss. This is because, as seen in Fig. \ref{fig:PositiveWake}, the lower stagnation point in the subcritical case is far away from the rear end of the bubble. Hence, the bubbles just coalesce with each other on contact. 

\subsubsection{Supercritical bubble pair interaction}

A strikingly different interaction is observed for the supercritical bubble pair. The snapshots of the interaction between the two supercritical bubbles rising inline in VE1 fluid are shown in Fig. \ref{fig:ViscoelasticInteraction} for a $Re$ $\approx$ 4 and Wi = 240 (see Supplemental Material). Similar to the subcritical case, in the supercritical bubble pair, the leading bubble’s wake region attracts the trailing bubble and thus the drafting-kissing phase is observed. Following that, instead of the expected tumbling or coalescence, the trailing bubble overtakes the leading bubble. This interchanging of the relative leading and trailing positions between the supercritical bubbles is referred to as `dancing’. This dancing process continues throughout the interaction as the bubble pair rises to the free surface.  In the supercritical bubbles, as the elastic forces are dominant, the bubble coalescence is delayed \cite{dekee1986bubble}. 

\begin{figure}[H]
\centering
    \includegraphics[scale=0.7]{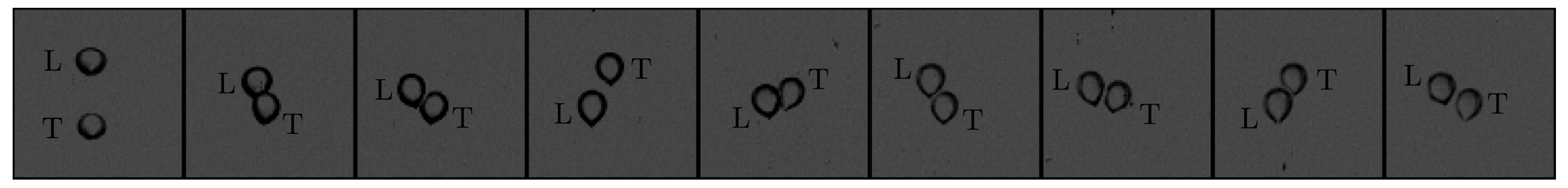}
\caption{Snapshots of images showing the interaction between a bubble pair (with a diameter of 5 mm) rising inline in the VE1 fluid at $Re$ $\approx$ 4 and Wi = 240. No tumbling or coalescence is observed. Instead, the two bubbles interchange their leading and trailing positions as they rise towards the free surface. This process is referred to as the drafting-kissing-dancing (DKD). L and T denote the leading and trailing bubbles at the beginning of the experiment respectively.}
  \label{fig:ViscoelasticInteraction}
\end{figure}
 
\begin{figure}[H]
    \centering
    \includegraphics[scale=0.45]{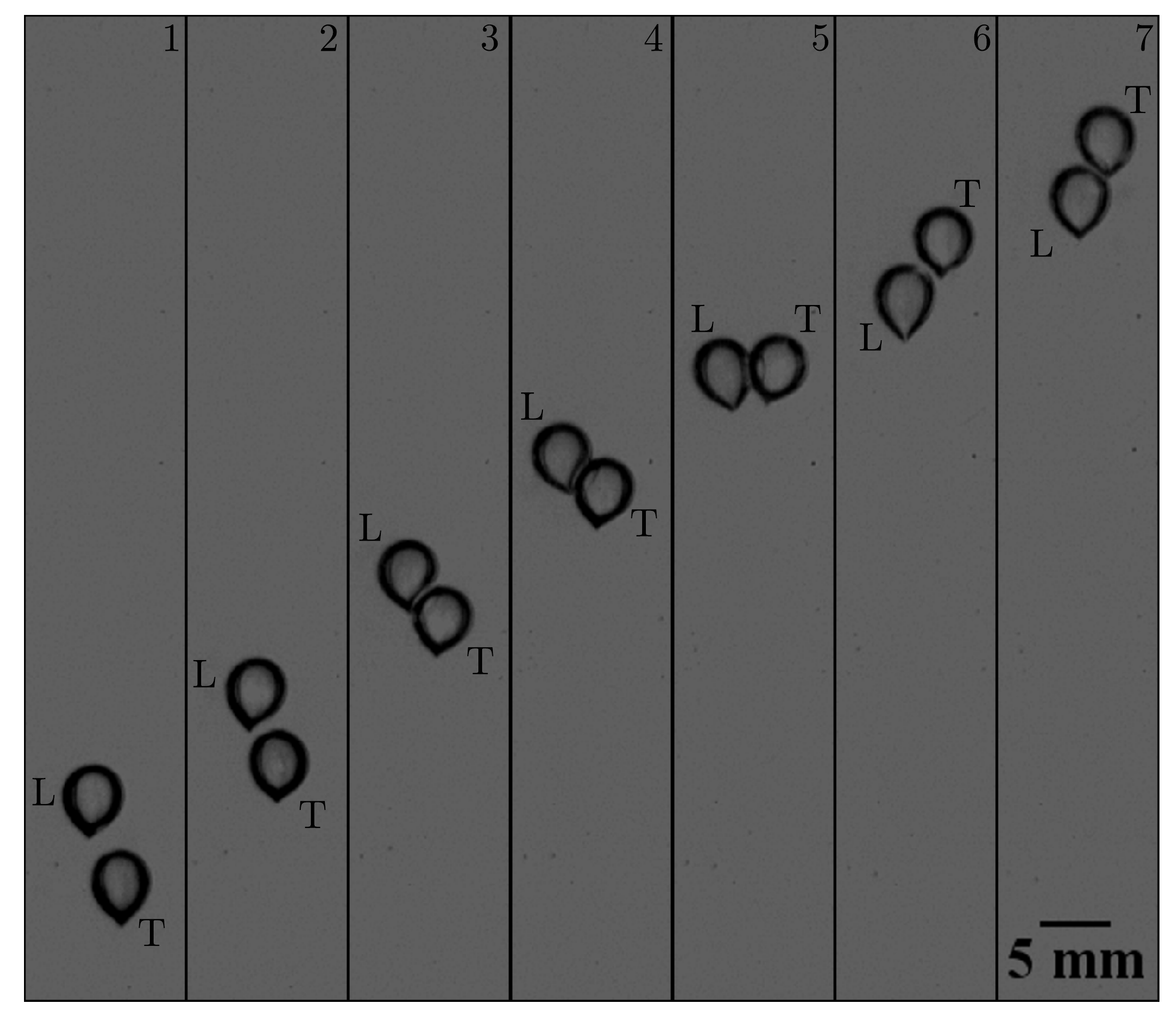}
    \caption{Series of images showing the dancing phase between two supercritical bubbles in the VE1 fluid. Interval between the images is 75 millisecond. Significant change in the shape can be observed when the bubbles interchange their relative positions. The trailing bubble retains its shape. However, the leading bubble deforms a lot and loses its pointed cusp shape. This can be observed clearly by comparing the 1st and the 5th frames. L and T denote the leading and trailing bubbles at the beginning of the experiment respectively.}
    \label{fig:ViscoelasticInteraction-Detailed}
\end{figure}

Since the bubbles interchange their positions, to be consistent with the nomenclature, the leading and trailing bubbles at the start of the experiment will always be identified as the leading and trailing bubble irrespective of their instantaneous transient positions. In this way, the changes experienced by the individual bubbles can be studied. On close observations,  as seen in Fig. \ref{fig:ViscoelasticInteraction-Detailed}, the leading bubble aligns itself off-centered from the trailing bubble even before the kissing phase (Frames 1-2). Following that, the leading bubble deforms to an inverted teardrop shape (Frames 5-6). 

\begin{figure}[H]
    \begin{subfigure}[h]{0.51\textwidth}
    \centering
    \includegraphics[scale=0.5]{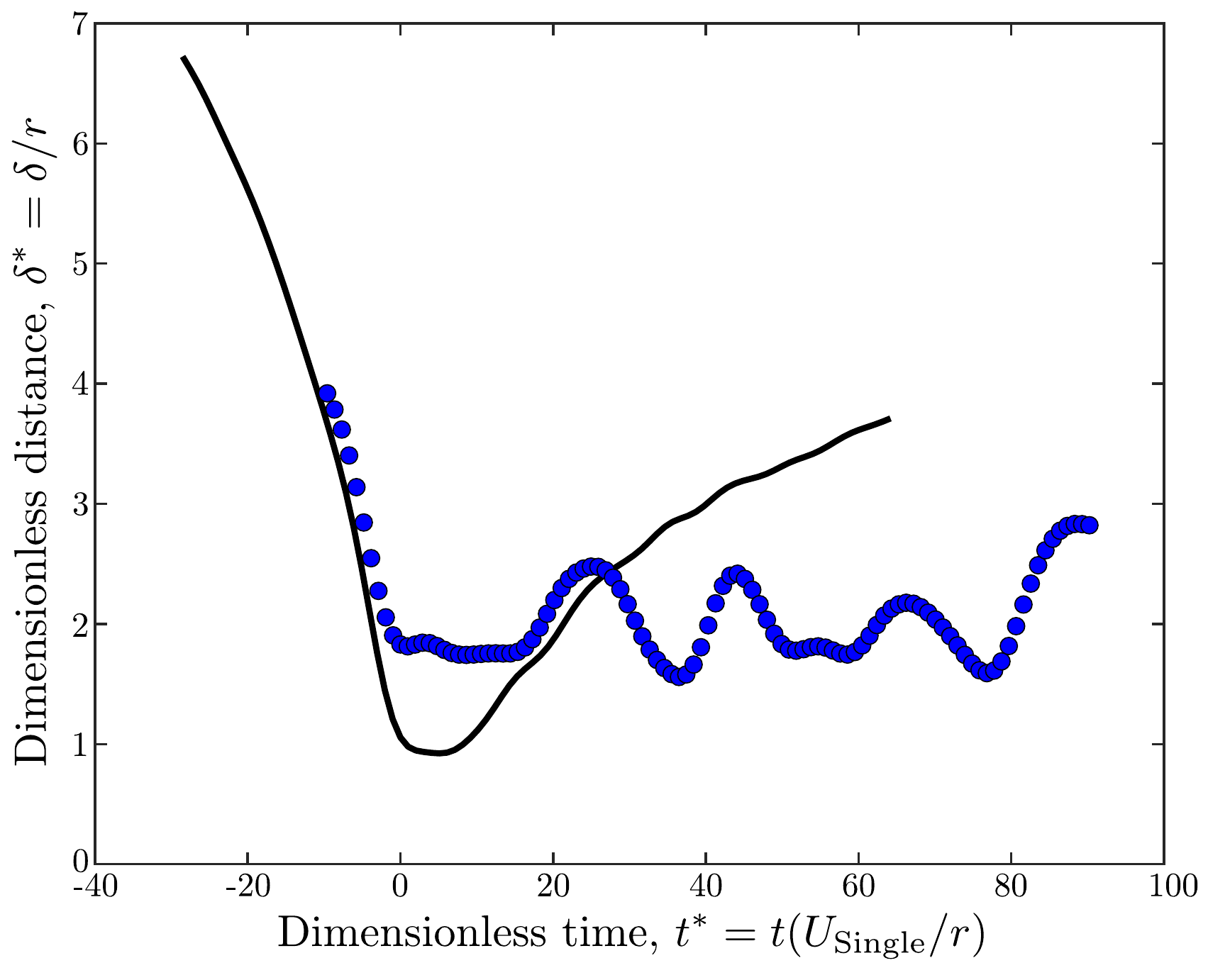}
    \caption{}
    \label{fig:ViscoelasticCharacteristic}
\end{subfigure}
\begin{subfigure}[h]{0.51\textwidth}
\centering
    \includegraphics[scale=0.5]{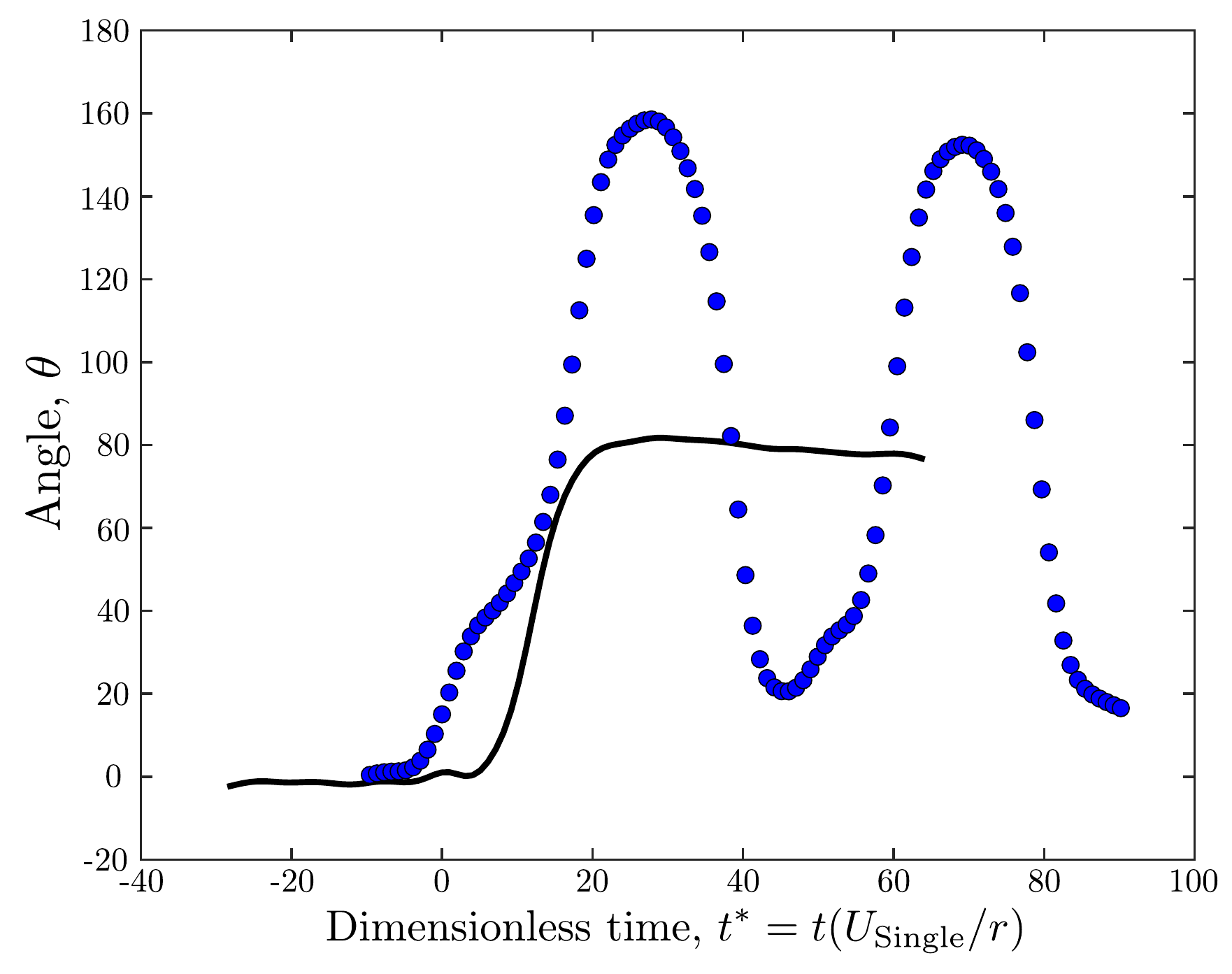}
    \caption{}
    \label{fig:ViscoelasticAngle}
\end{subfigure}
    \caption{Supercritical bubbles in the VE1 fluid ( \tikzcircle[fill=blue]{4pt} ): (\ref{fig:ViscoelasticCharacteristic}) Dimensionless distance between the two bubbles rising as a function of dimensionless time; (\ref{fig:ViscoelasticAngle}) Angle made by the line joining the bubble centers with respect to the vertical axis as a function of dimensionless time. The Reynolds number and Weissenberg number are 4 and 240 respectively. The results are compared to that of the classical DKT process observed for the two bubbles in the Newtonian fluid (---). }
    \label{fig:Viscoelastic}
\end{figure}

Fig. \ref{fig:ViscoelasticCharacteristic} shows the dimensionless distance between two bubbles, $\delta^{*} = \delta/r$, as a function of dimensionless time, $t^{*} = t(U_{\text{Single}}/r)$, in the VE1 fluid. Note that $t^{*}$ = 0 is selected to be the time at which the bubbles first ``touch" each other (smallest value of $\delta^{*}$). The dimensionless distance  between the supercritical bubbles oscillates between a peak and a trough during the dancing phase. Though the separation distance between the two bubbles exceeds about 2.5 times the bubble radius, the bubble in the trailing position still catch up with that in the leading position. Experiments conducted by varying the separation distance between the bubbles (not shown here) did not reveal a significant change in the interaction. Fig. \ref{fig:ViscoelasticAngle} depicts the change in angle made by the line joining the bubble centers with respect to the vertical axis.

\begin{figure}[H]
    \begin{subfigure}[h]{0.51\textwidth}
    \centering
    \includegraphics[scale=0.5]{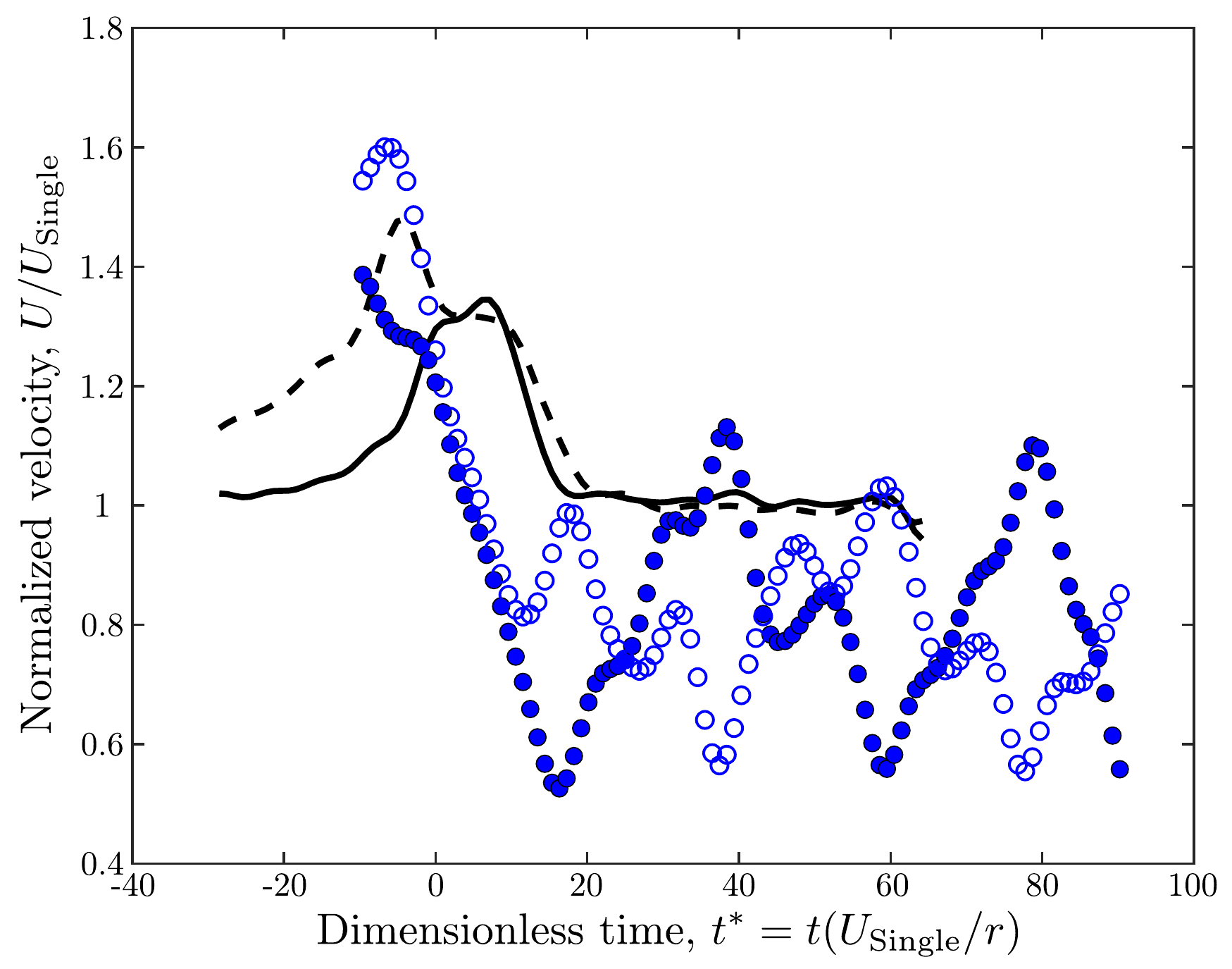}
    \caption{}
    \label{fig:ViscoelasticVelocity}
\end{subfigure}
\begin{subfigure}[h]{0.51\textwidth}
\centering
    \includegraphics[scale=0.5]{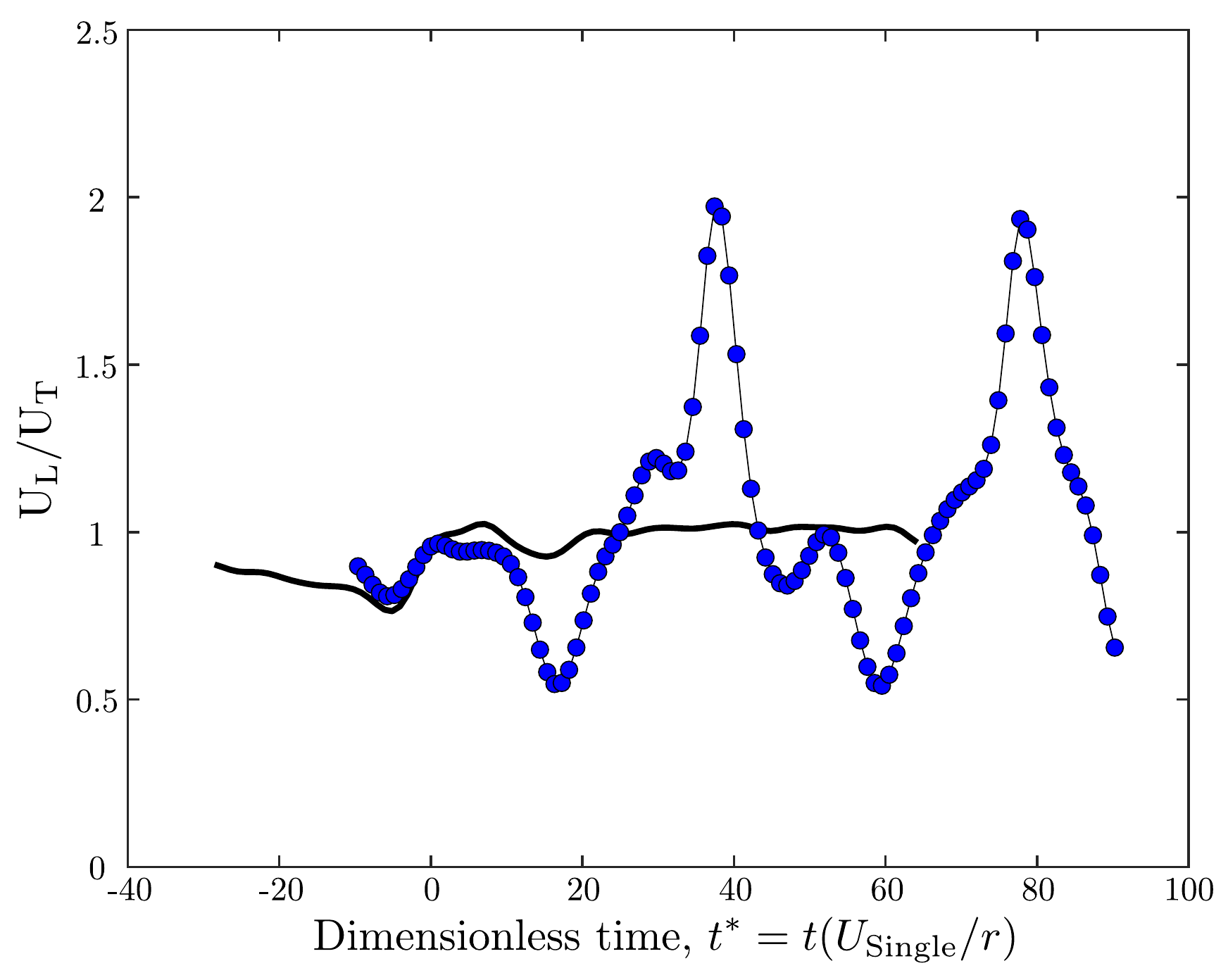}
    \caption{}
    \label{fig:ViscoelasticCDTL}
\end{subfigure}
    \caption{Same data as in Fig. \ref{fig:Viscoelastic}. Supercritical bubbles in the VE1 fluid: (\ref{fig:ViscoelasticVelocity}) Velocity of the leading bubble ( \tikzcircle[fill=blue]{4pt} ) and trailing bubble  ( \tikzcircle[draw=blue]{4pt} ) normalized by the single bubble velocity as a function of dimensionless time. This is compared to the velocity of the leading bubble (---) and trailing bubble (- -) in the Newtonian fluid;  (\ref{fig:ViscoelasticCDTL})  Velocity ratio ( \tikzcircle[fill=blue]{4pt} ) of the leading bubble (U$_{\text{L}}$) to the trailing bubble (U$_{\text{T}}$) as a function of dimensionless time. This is compared to that of the bubbles in the Newtonian fluid (---).
}
    \label{fig:Viscoelastic0.15}
\end{figure}

From the normalized velocity plot, Fig. \ref{fig:ViscoelasticVelocity}, it can be seen that during the kissing phase, the velocities of the bubbles as a system decreases.  In the dancing phase, however, the velocities of the leading and trailing bubbles are always out of phase with each other. Respect to their relative positions, the velocities of the leading and trailing bubbles rise or drop proportionately. This can be observed by plotting the ratio of instantaneous velocities of the leading bubble ($\text{U}_\text{L}$) to that of the trailing bubble ($\text{U}_\text{T}$) as a function of dimensionless time in Fig. \ref{fig:ViscoelasticCDTL}.

\begin{figure}[H]
\centering
\subfloat[t* = -0.6 ]{\label{ACV_a}\includegraphics[scale=0.25]{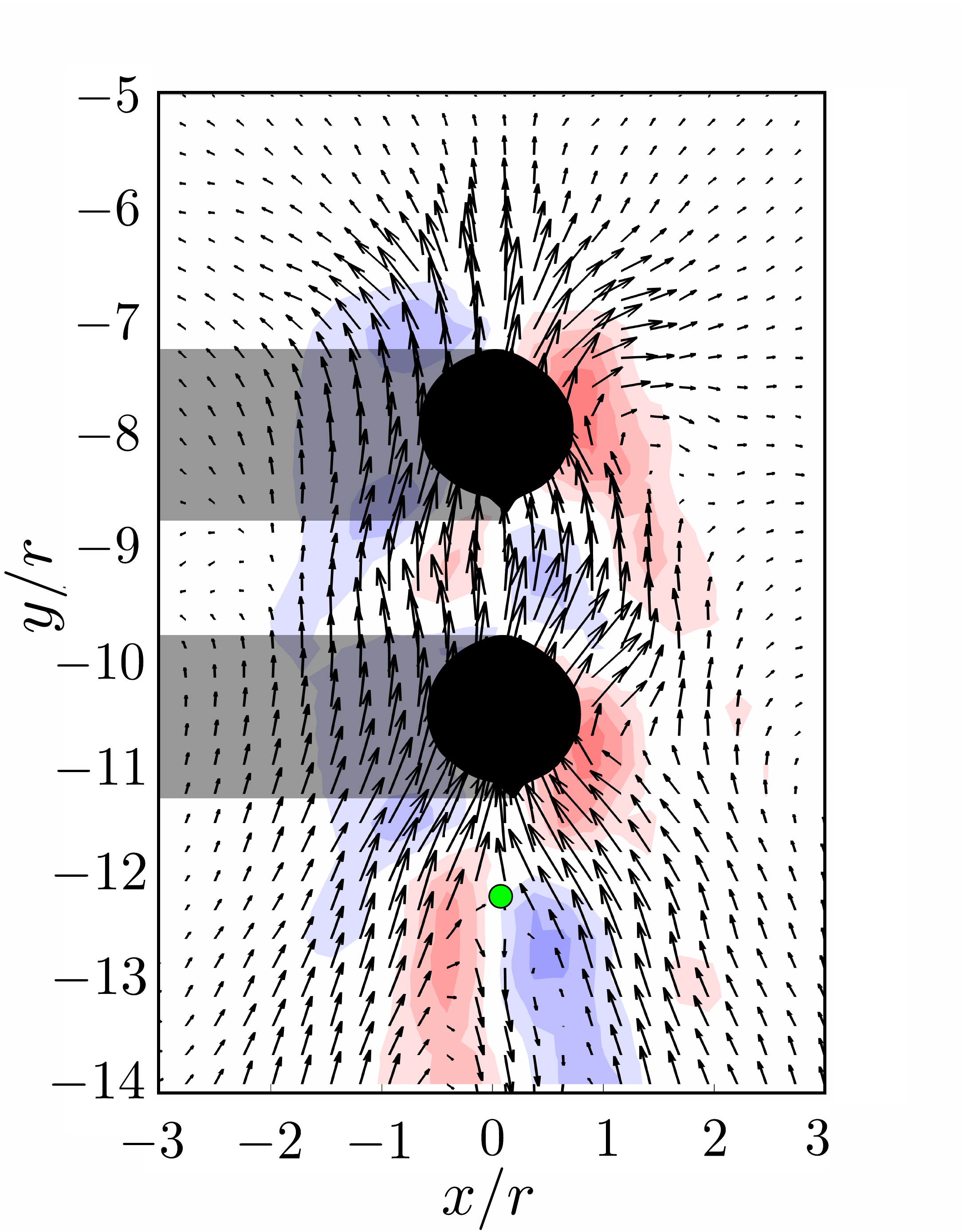}}
\subfloat[t* = 0 ]{\label{ACV_b}\includegraphics[scale=0.25]{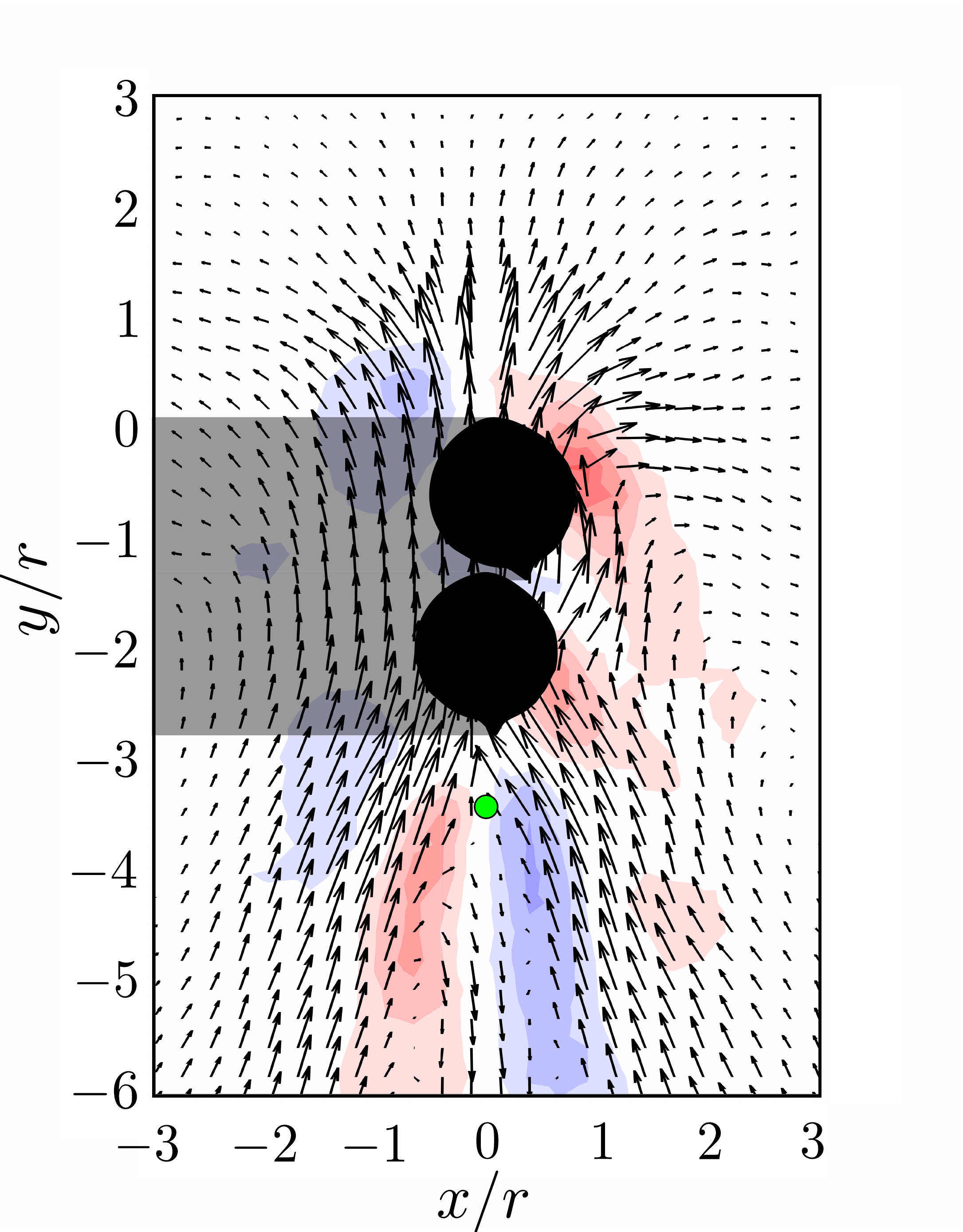}}
\subfloat[t* = 0.6 ]{\label{ACV_c}\includegraphics[scale=0.25]{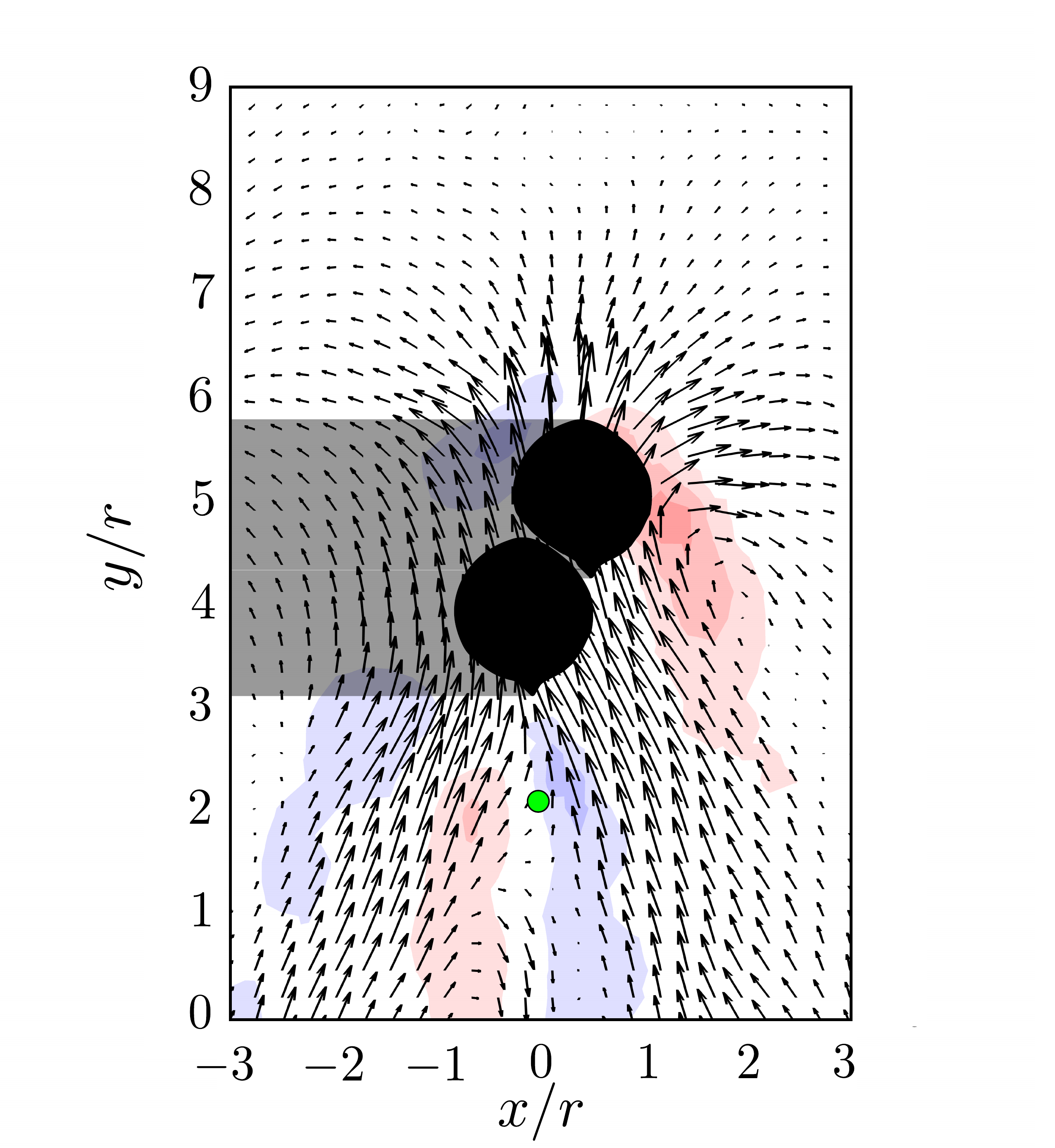}}\hfill
\subfloat[t* = 7.2 ]{\label{ACV_d}\includegraphics[scale=0.25]{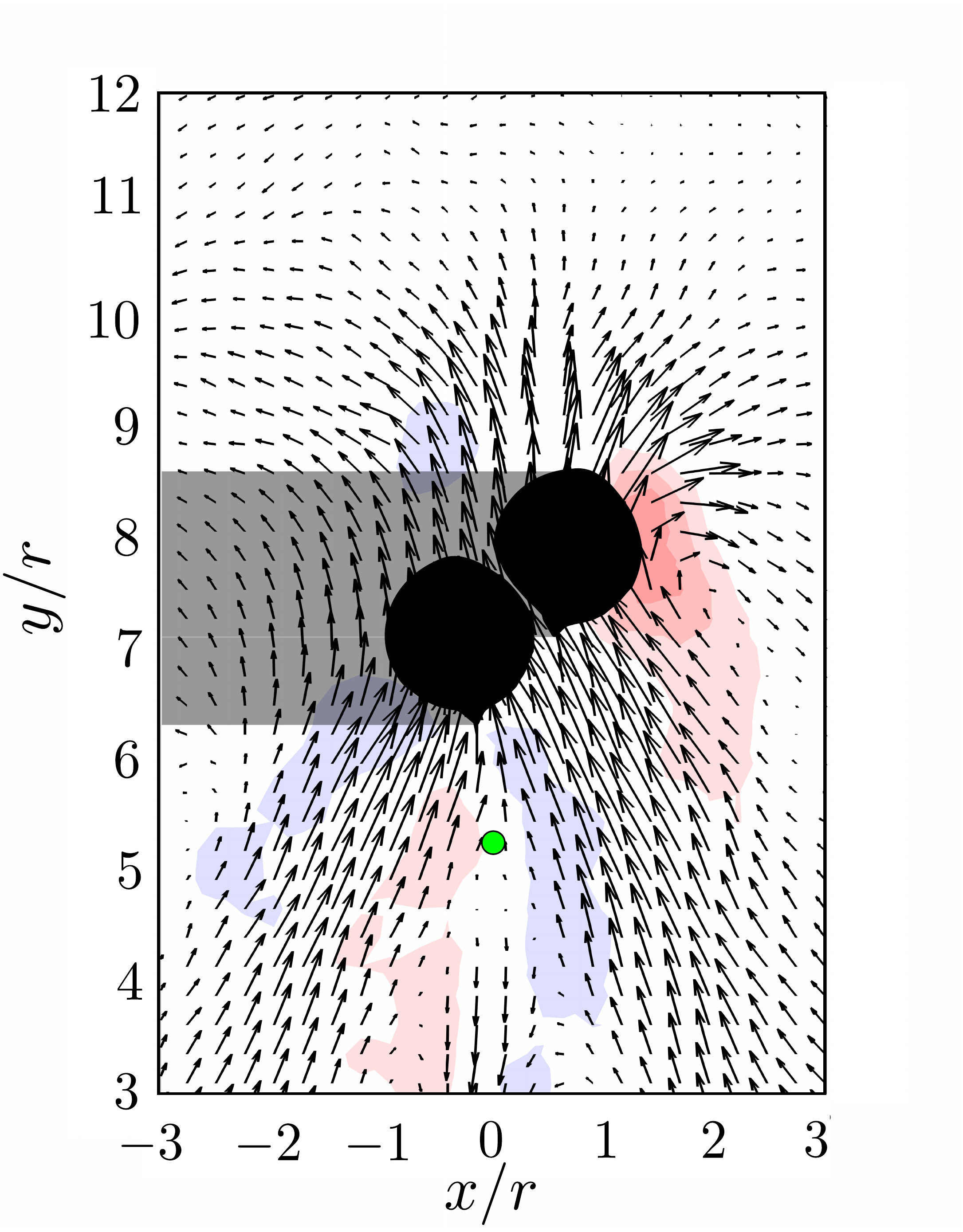}}
\subfloat[t* = 9.6 ]{\label{ACV_e}\includegraphics[scale=0.25]{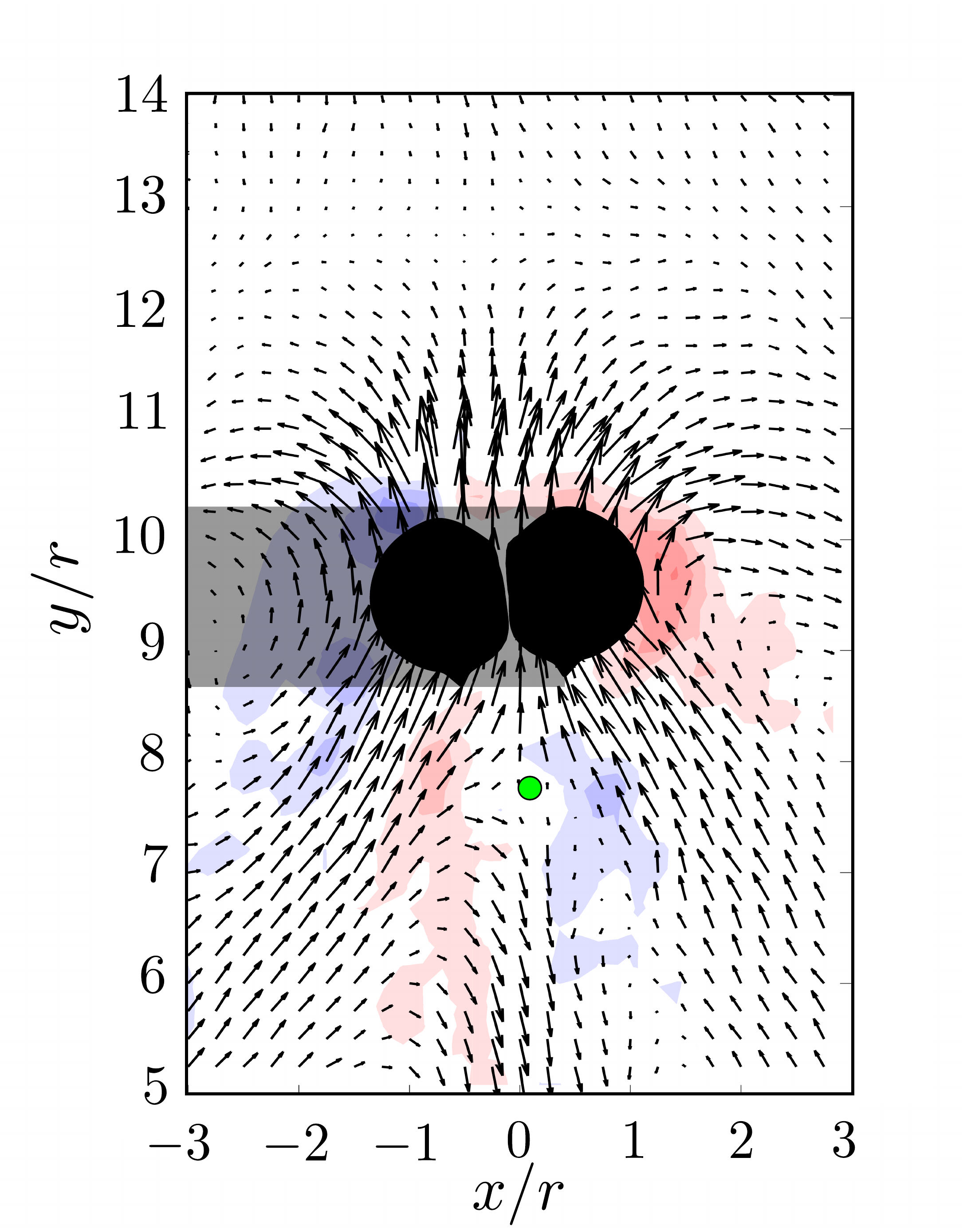}}
\subfloat[t* = 14.4 ]{\label{ACV_f}\includegraphics[scale=0.25]{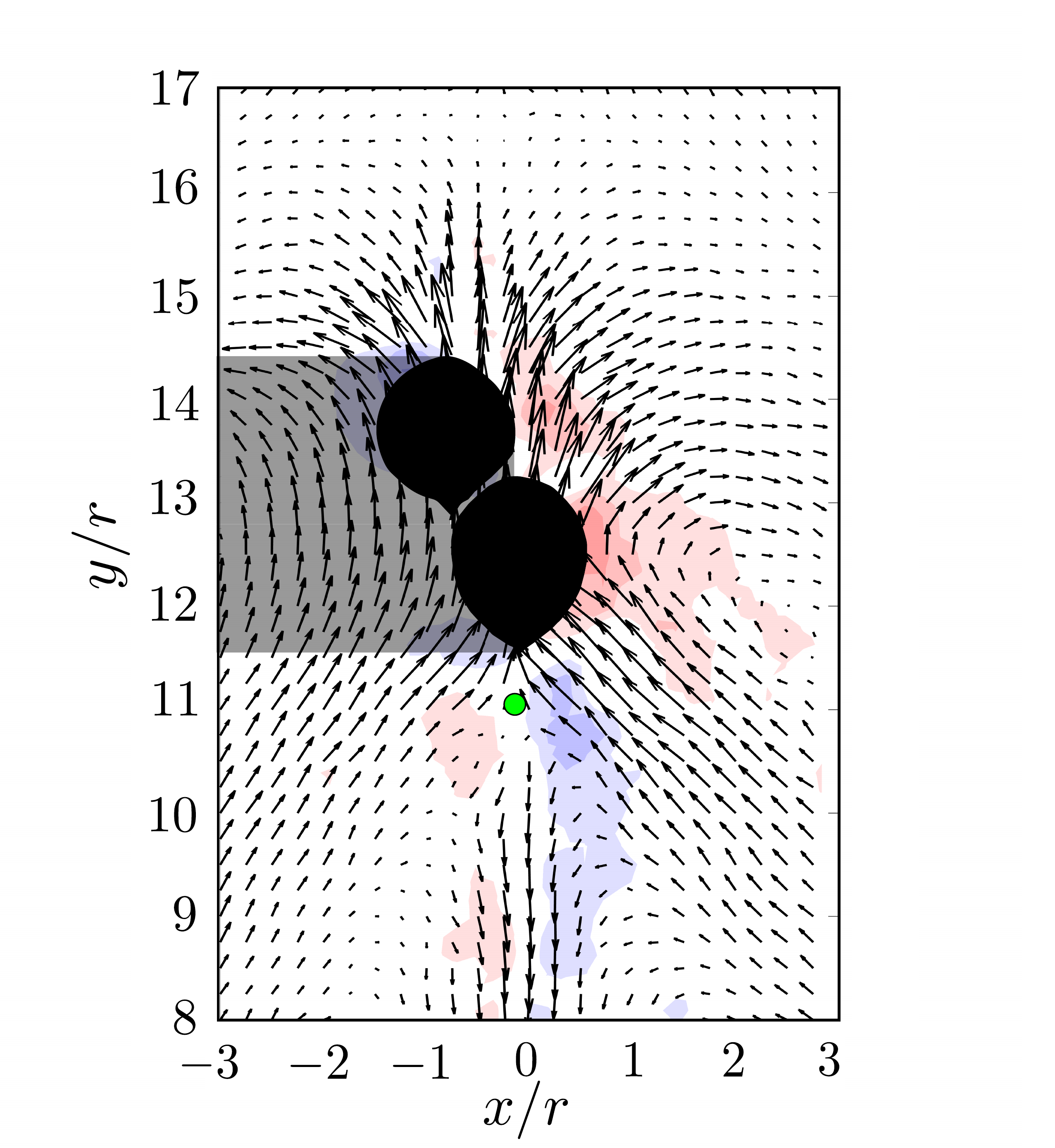}}\hfill
\subfloat[t* = 17.6 ]{\label{ACV_g}\includegraphics[scale=0.25]{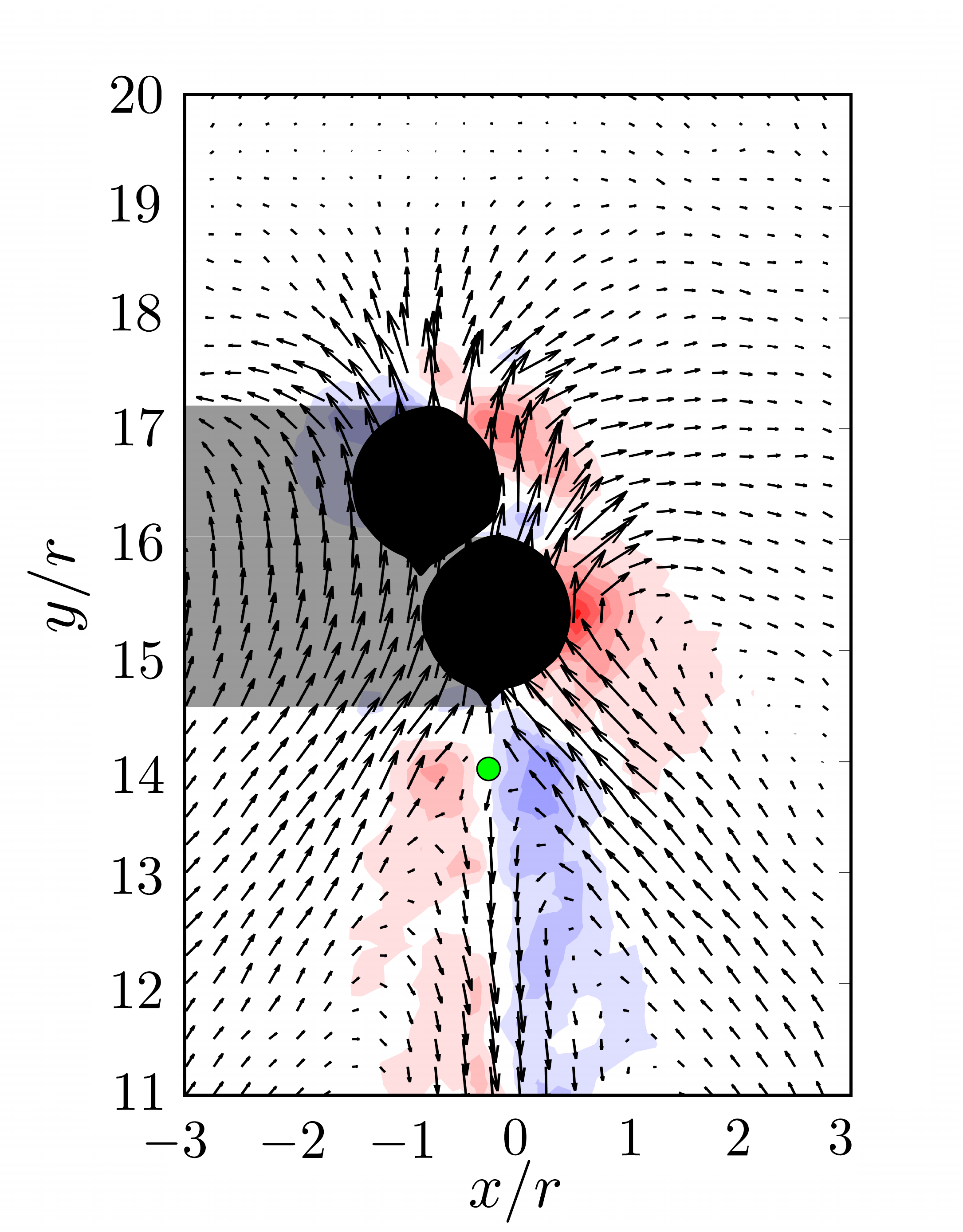}}
\subfloat[t* = 20.8 ]{\label{ACV_h}\includegraphics[scale=0.25]{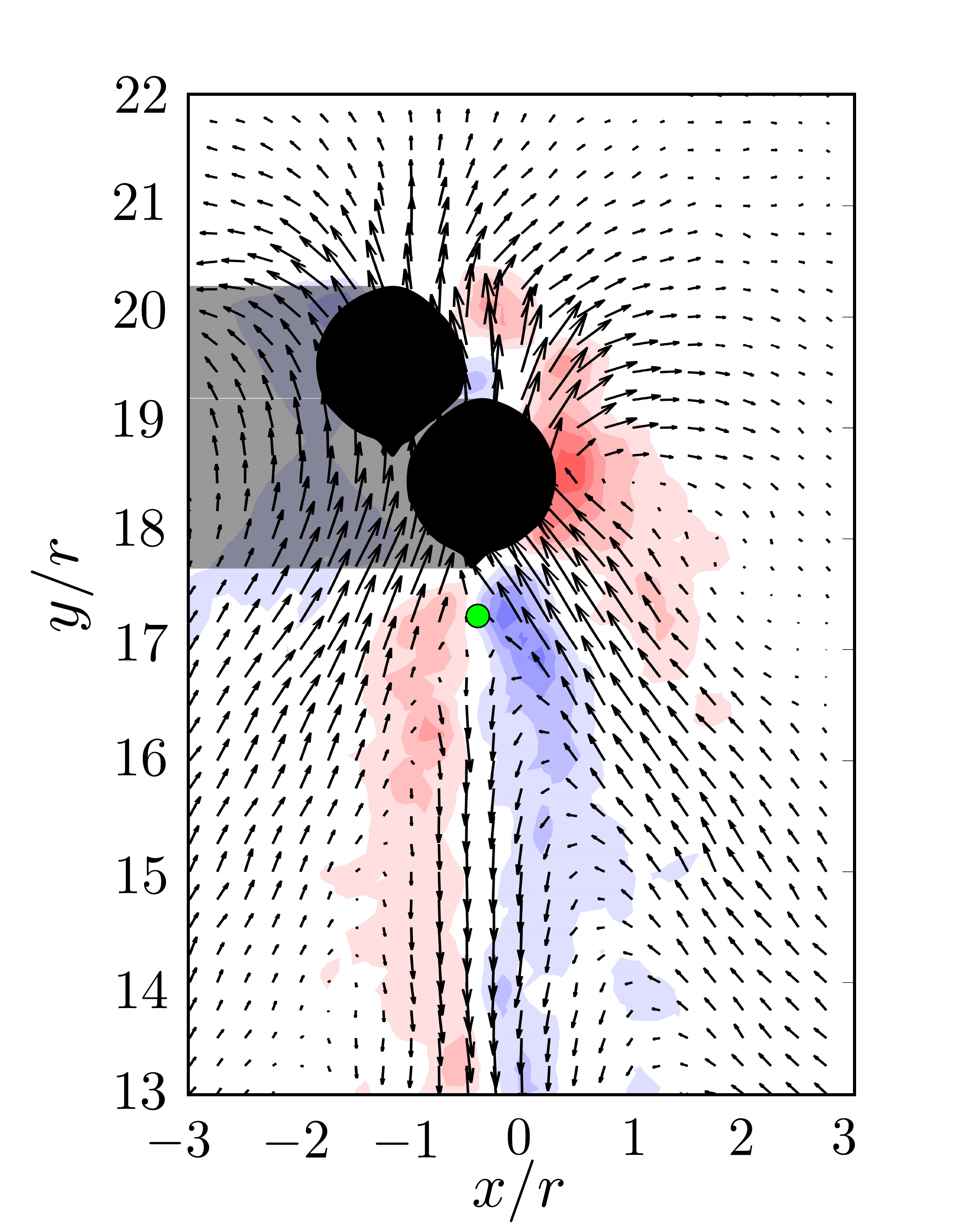}}
\subfloat[t* = 24 ]{\label{ACV_i}\includegraphics[scale=0.25]{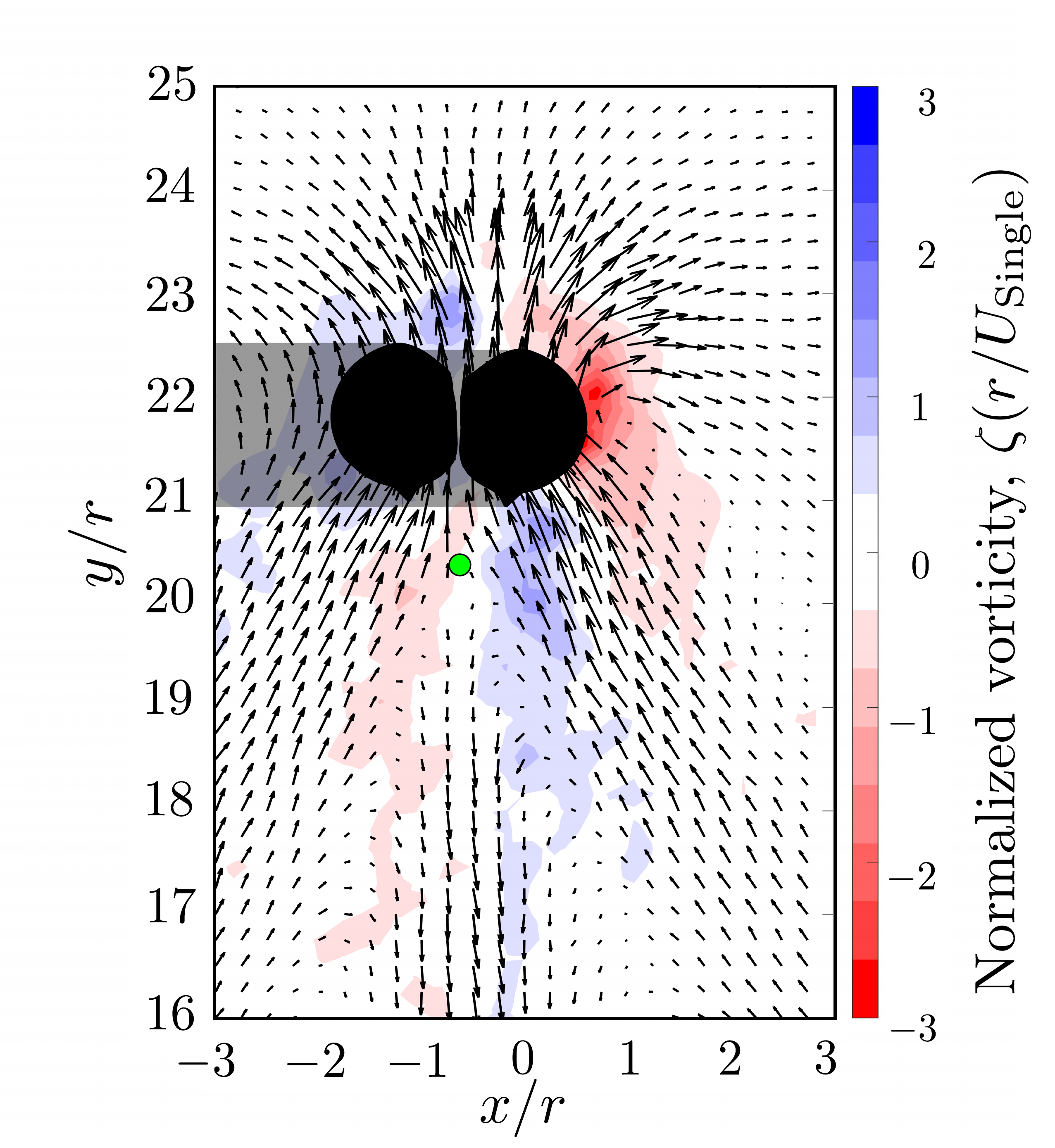}}\hfill\par
\caption{Instantaneous velocity fields around the supercritical bubble pair rising inline in the VE1 fluid. The Reynolds number and Weissenberg number are 4.5 and 200, respectively. The position at which the bubbles first touch each other is considered to be at (0, 0) and the coordinates are normalized by the bubble radius. The contours depict the normalized vorticity fields around the bubble. The stagnation point in the wake of the bubble is indicated by a green dot.}
\label{fig:ACV_PIV}
\end{figure}

The results from the flow field visualization of the supercritical bubble pair interaction over the dimensionless time is shown in the Fig. \ref{fig:ACV_PIV}. Here, when the trailing bubble approaches the leading bubble, the flow field behind the leading bubble is strongly modified and no negative wake was observed behind the leading bubble (Fig. \ref{ACV_a}). However, the sign of vortices in between the bubbles is opposite to that of the vortices near the bubble's equator. This is an important observation because in the supercritical bubbles, the lower stagnation point is very close to the rear side of the bubble, almost one bubble radius distance away (Fig. \ref{fig:NegativeWake}). Thus the leading bubble experiences the presence of the trailing bubble even before they touch. This explains why the supercritical bubble pair do not coalesce with each other as in the case of subcritical bubbles.  This flow field modification is similar to that of the flow fields around two sedimenting spheres in the viscoelastic shear-thinning fluids reported by Verneuli et al. \cite{verneuil2007axisymmetric}. From the complimentary experiments conducted with solid spheres at the same Reynolds number (see Appendix), we observed the well reported formation of stable vertically aligned sphere pairs \cite{joseph1994aggregation,verneuil2007axisymmetric}. However, unlike the solid spheres, the supercritical bubble pair does not align themselves vertically. Instead, the bubble pair adopt a diagonal alignment (Fig. \ref{ACV_b} and \ref{ACV_c}) similar to the case in the shear-thinning inelastic fluids reported by Velez-Cordero et al. \cite{velez2011hydrodynamic}. This is an indication that the bubble deformability influence the interaction observed in the supercritical bubble pair.  

Now, in the diagonal alignment, the trailing bubble accelerates in the reduced viscosity path of the leading bubble, and then rise past the leading bubble (Fig. \ref{ACV_d}, \ref{ACV_e} and \ref{ACV_f}). This process continues as the bubble pair rises to the free surface (Fig. \ref{ACV_g}, \ref{ACV_h} and \ref{ACV_i}). As mentioned earlier, the shape of the leading bubble is deformed to an inverted teardrop when the trailing bubble overtakes the leading (Fig. \ref{ACV_f}). When the leading bubble regains its original shape, its velocity increases and the bubble jumps forward (Fig. \ref{ACV_g}). A similar behavior on the transient oscillatory motion of ascending bubbles due to the shape change was reported by Handzy et al. \cite{handzy2004oscillatory} in wormlike micellar fluids. This might also be due to the high polymeric stresses around the trailing bubble reported by Yuan et al. \cite{yuan2021vertical} on the vertical chain of bubbles. Furthermore, in their concluding remarks, Bothe et al. \cite{bothe2022molecular} posed a thought experiment, that is, any disturbance to the rise velocity of a bubble in viscoelastic fluid, at which the relaxation induced effects are balanced, will induce either a rise or fall to the velocity. Our experimental results suggest that this rise or fall in the bubble velocity is achievable with the presence of a trailing bubble in the leading bubble's wake. 

\subsection{Effect of polymer concentration on the bubble-bubble interaction}

\begin{figure}[H]
    \begin{subfigure}[h]{0.51\textwidth}
    \centering
    \includegraphics[scale=0.5]{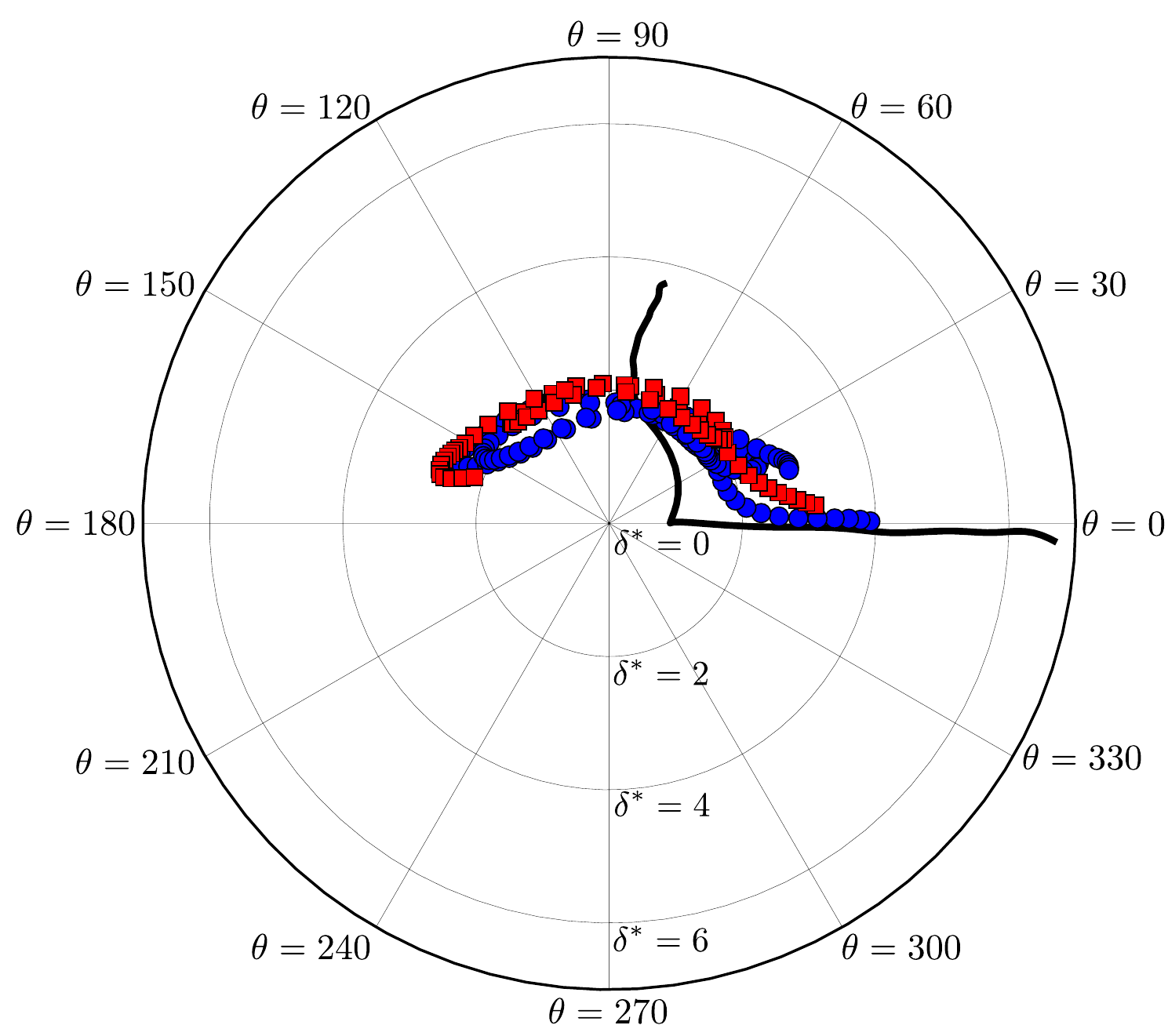}
    \caption{}
    \label{fig:PolymerPolar}
\end{subfigure}
\begin{subfigure}[h]{0.51\textwidth}
\centering
    \includegraphics[scale=0.5]{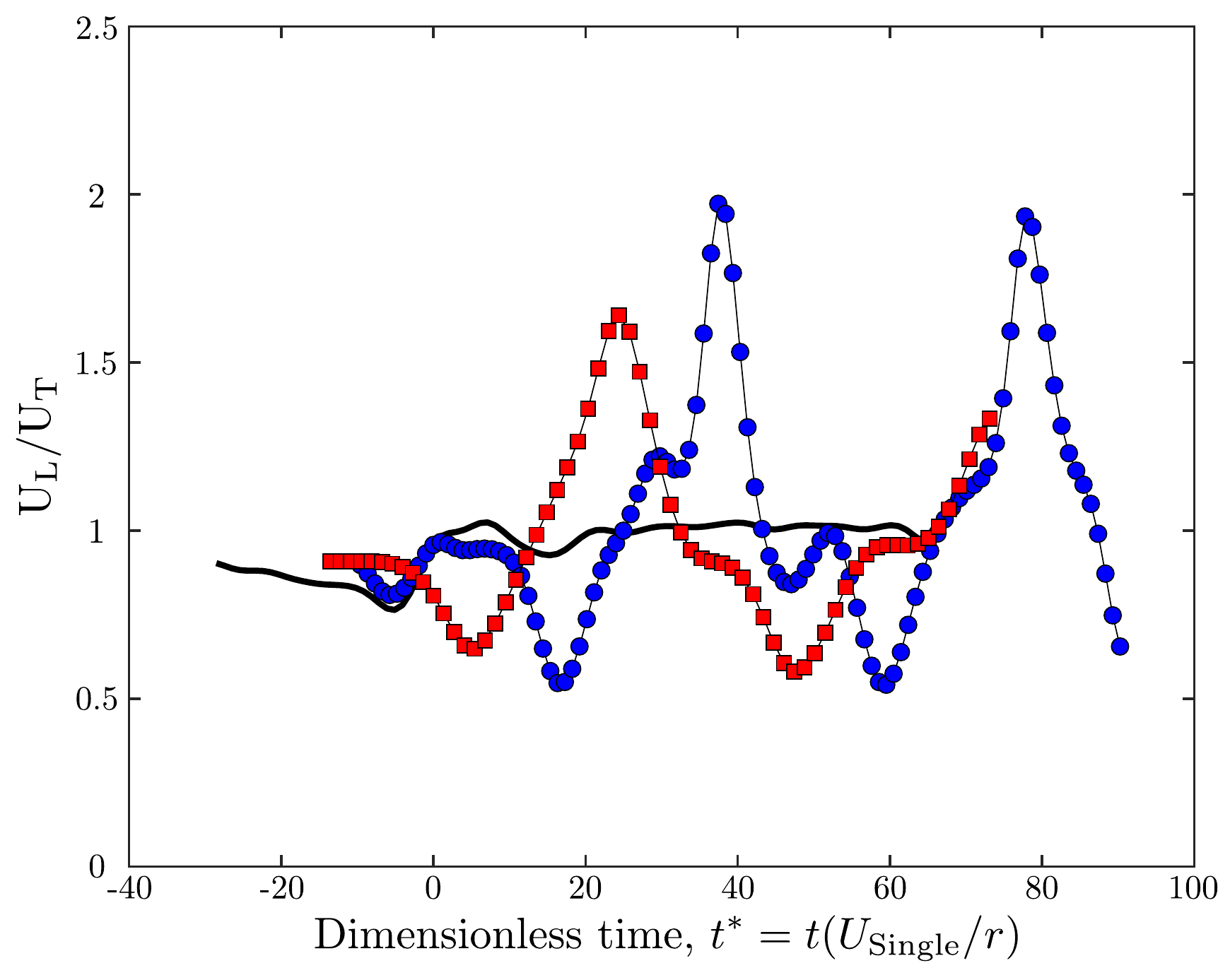}
    \caption{}
    \label{fig:PolymerCDTL}
\end{subfigure}
     \caption{Supercritical bubbles in the VE2 fluid ( \tikzsquare[fill=red]{10pt}) and in the  VE1 fluid ( \tikzcircle[fill=blue]{4pt} ): (\ref{fig:PolymerPolar}) Dimensionless distance between the two bubbles as a function of angle in polar plots;  (\ref{fig:PolymerCDTL}) Velocity ratio of the leading bubble (U$_{\text{L}}$) to the trailing bubble (U$_{\text{T}}$) as a function of dimensionless time. The results are compared to that of the classical DKT process observed for the two bubbles in the Newtonian fluid (---). }
\end{figure}

To understand the effect of polymer concentration on the DKD process, bubble-bubble interaction was studied in the VE2 fluid (see Table \ref{table:FluidProperties}). To have a comparable Reynolds number, an equivalent bubble diameter of 6.2 mm was selected. For simplicity, the results are presented in the polar plots. In the Fig. \ref{fig:PolymerPolar}, the concentric circles correspond to dimensionless distance and the radial lines correspond to the angle made by the line joining the bubble centers with respect to the vertical axis. 

Excluding the dependence on time, from Fig. \ref{fig:PolymerPolar}, it can be seen that the dimensionless distance between the bubbles in the VE2 fluid stays almost constant after the kissing phase for $Re$ $\approx$ 4 and Wi = 850. Furthermore, the amplitude of the angle between the bubbles in the VE2 fluid is less than that of in the VE1 fluid. This implies that the tendency of the bubbles to stay close together is higher for the higher polymer concentration. In Fig. \ref{fig:PolymerCDTL}, it is evident that the velocity ratio of the leading to the trailing bubble in the VE2 fluid is less than that in the VE1 fluid. Though, this agrees with the expectation that the bubbles are more stable as the elasticity of the fluid increases \cite{yuan2021vertical}, it is not yet possible to conclude that the elasticity stabilizes the bubble oscillations. The distinct DKD interaction of bubbles in viscoelastic shear-thinning fluids can be a synergistic effect of both the shear-dependent viscous and elastic forces. This will be discussed in detail in section 3.5.

\subsection{Effect of salt on the bubble-bubble interaction}

\begin{figure}[H]
    \begin{subfigure}[h]{0.51\textwidth}
    \centering
    \includegraphics[scale=0.5]{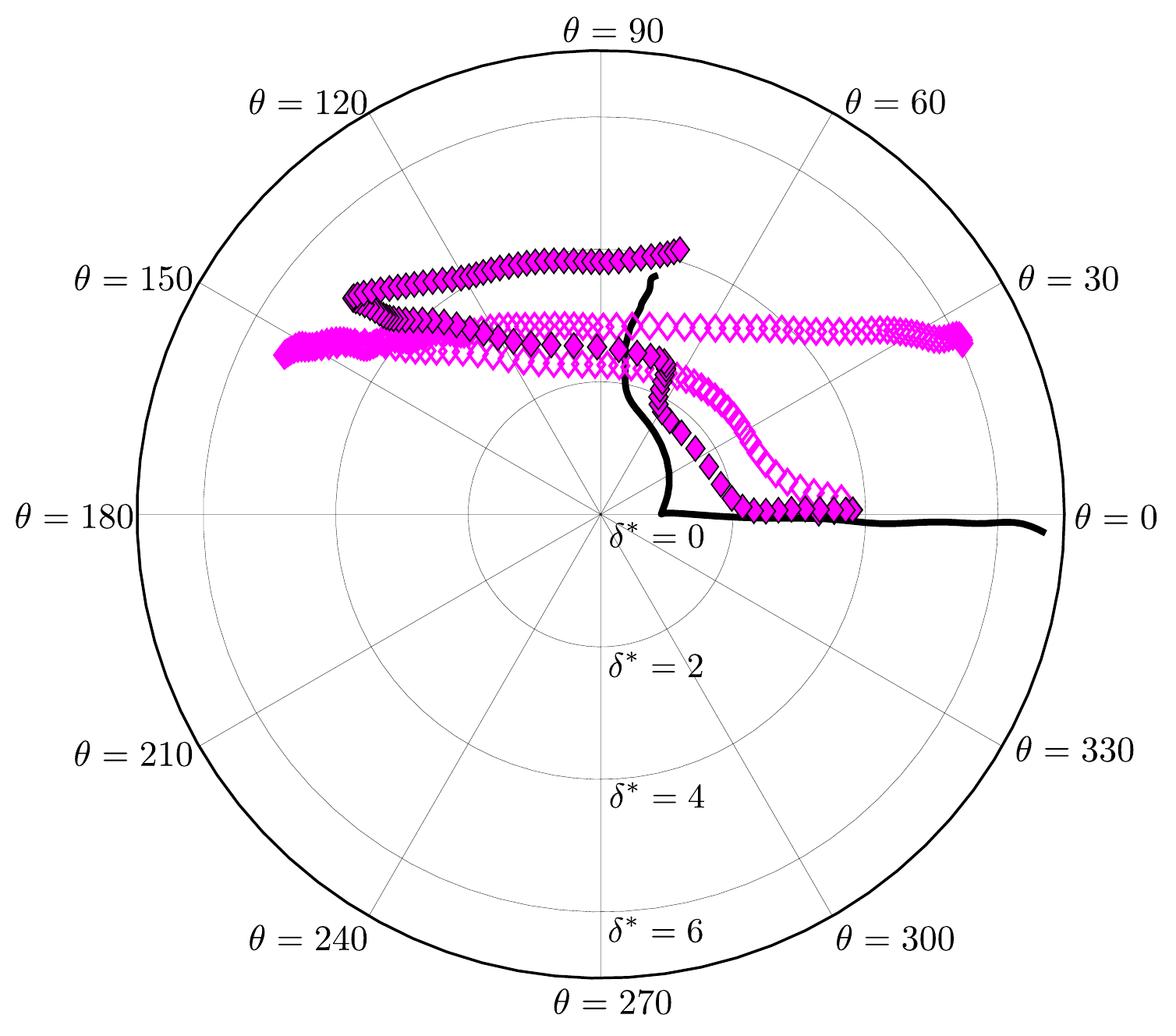}
    \caption{}
    \label{fig:SaltPolar}
\end{subfigure}
\begin{subfigure}[h]{0.51\textwidth}
\centering
    \includegraphics[scale=0.5]{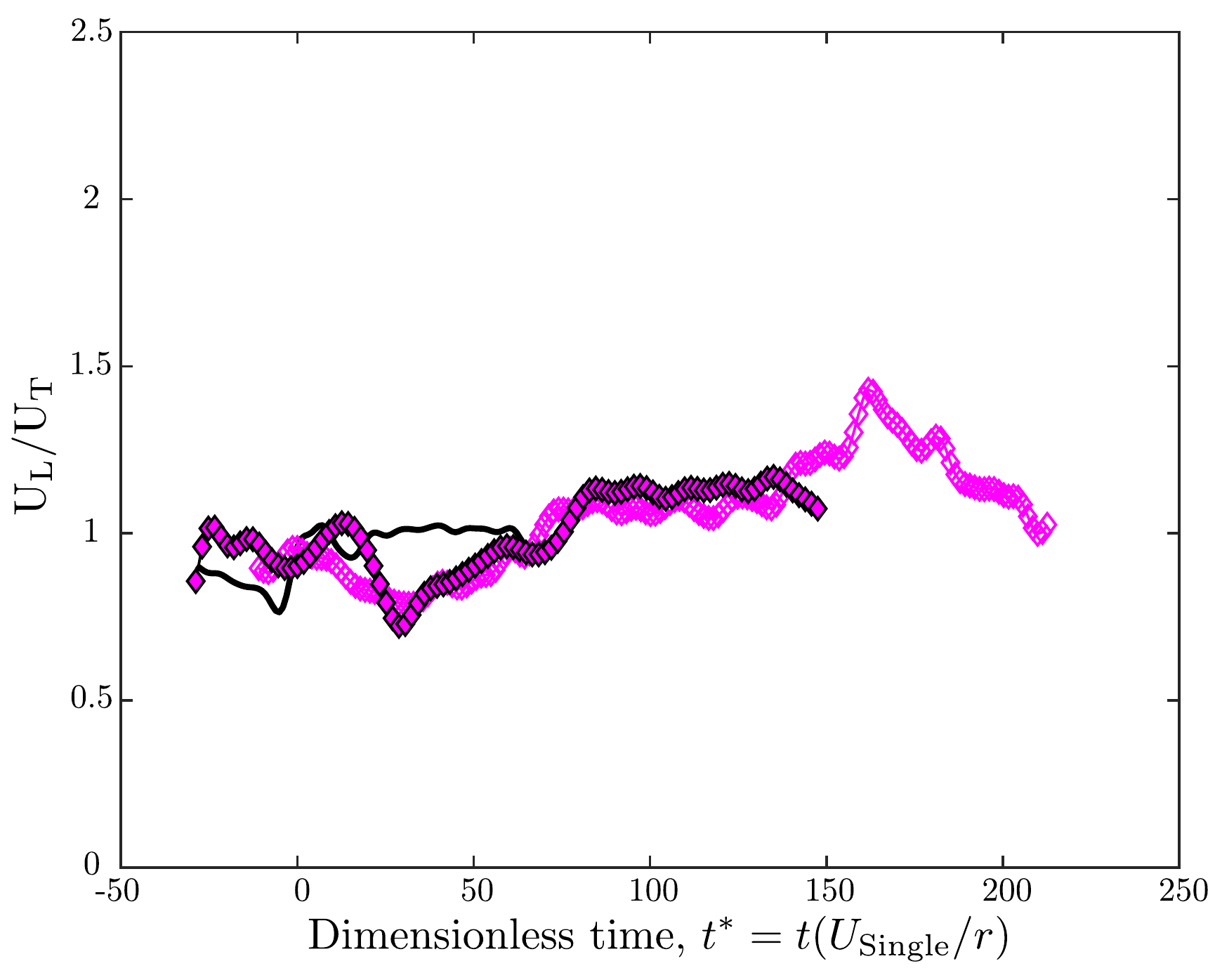}
    \caption{}
    \label{fig:SaltCDTL}
\end{subfigure}
     \caption{The open symbol ( \tikzdiamond[draw=magenta]{10pt}) corresponds to subcritical bubble pair and the closed symbols ( \tikzdiamond[fill=magenta]{10pt}) corresponds to supercritical bubble pair in the VE1-S fluid: (\ref{fig:SaltPolar}) Dimensionless distance between the two bubbles as a function of angle in polar plots;  (\ref{fig:SaltCDTL}) Velocity ratio of the leading bubble (U$_{\text{L}}$) to the trailing bubble (U$_{\text{T}}$) as a function of dimensionless time. The results are compared to that of the classical DKT process observed for the two bubbles in the Newtonian fluid (---).}
\end{figure}

By decreasing the effective viscosity and achieving a comparable mean relaxation time, the effect of viscous forces on the DKD process can be studied. This was obtained by adding a small amount of MgSO$_4$ salt to the VE1 fluid, see VE1-S fluid in the Table \ref{table:FluidProperties}. Bubble-bubble interaction experiments were then repeated for subcritical and supercritical cases in the VE1-S fluid. Interestingly, for both the cases, the bubbles rising inline showed some resemblance to the DKD phenomenon. As seen in Fig. \ref{fig:SaltPolar}, the bubbles interchange their positions, but the dimensionless distance between them is larger. This could result from the short range repulsive force introduced by the addition of salt \cite{lessard1971bubble}. From Fig. \ref{fig:SaltCDTL}, it can be seen that the ratio of leading to the trailing bubble velocity follow a similar trend of rise and fall as observed in the case of supercritical bubbles in VE1 fluid.

The DKD process observed for the subcritical bubbles in VE1-S fluid for $Re$ $\approx$ 2 and Wi = 130 can be directly attributed to the increased elasticity of the fluid in the lower shear rate regime (Fig. \ref{fig:Oscillation}). However, the overall interaction is slower in the VE1-S fluid. One reason for this could be the decrease in the Reynolds number of the bubbles compared to that of the supercritical bubbles in VE1 fluid. Hence, the time required by the trailing bubble to catch up with the leading bubble is longer. In other words, the viscous forces speed up the interaction between the bubbles. This confirms that some balance between the viscous and the elastic forces should be achieved in order to successfully observe the DKD process in viscoelastic fluids.

\subsection{Comparison with the two-bubble interaction in shear-thinning inelastic fluid}

\begin{figure}[H]
    \begin{subfigure}[h]{0.51\textwidth}
    \centering
    \includegraphics[scale=0.5]{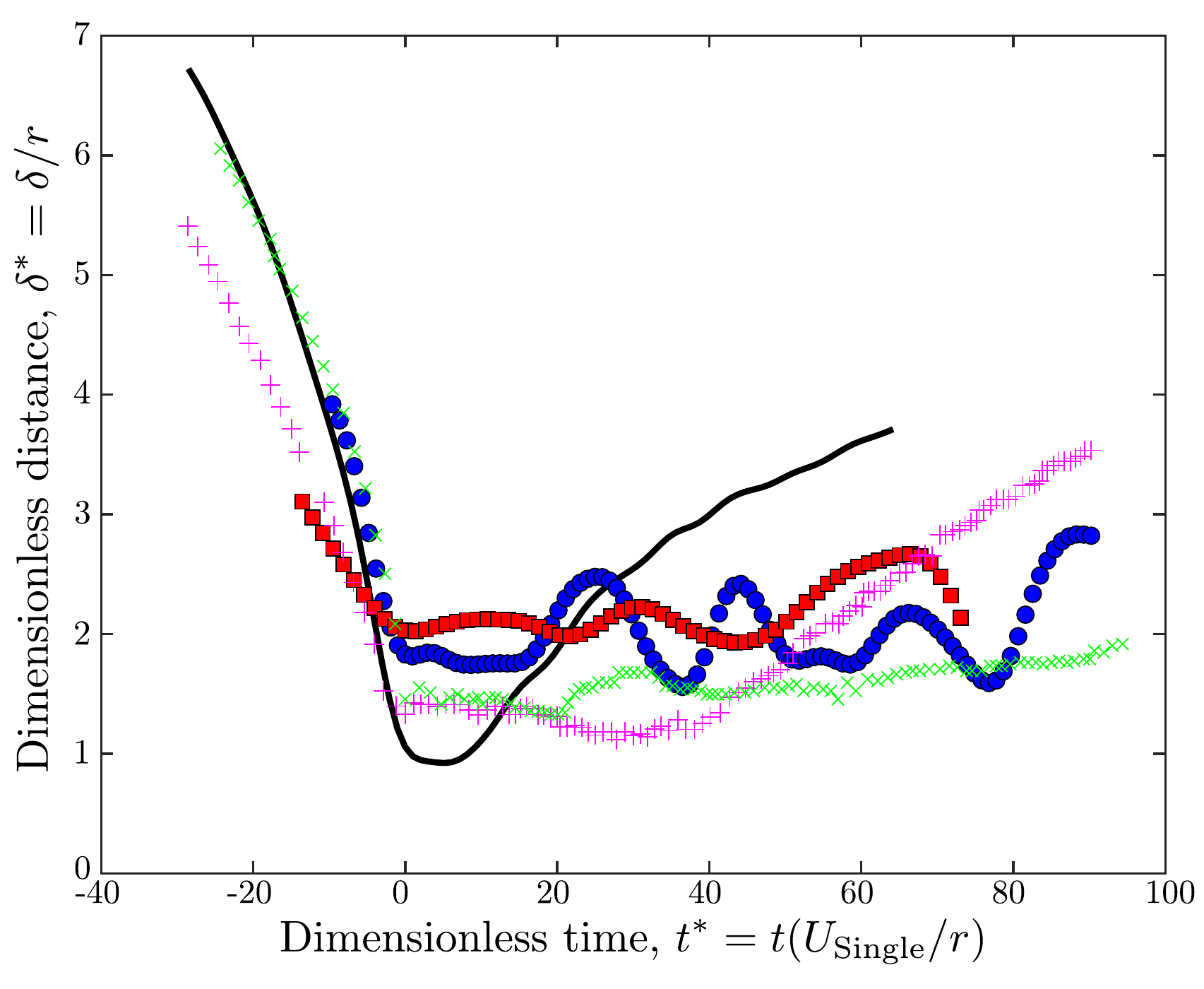}
    \caption{}
    \label{fig:STCharacteristic}
\end{subfigure}
\begin{subfigure}[h]{0.51\textwidth}
\centering
    \includegraphics[scale=0.5]{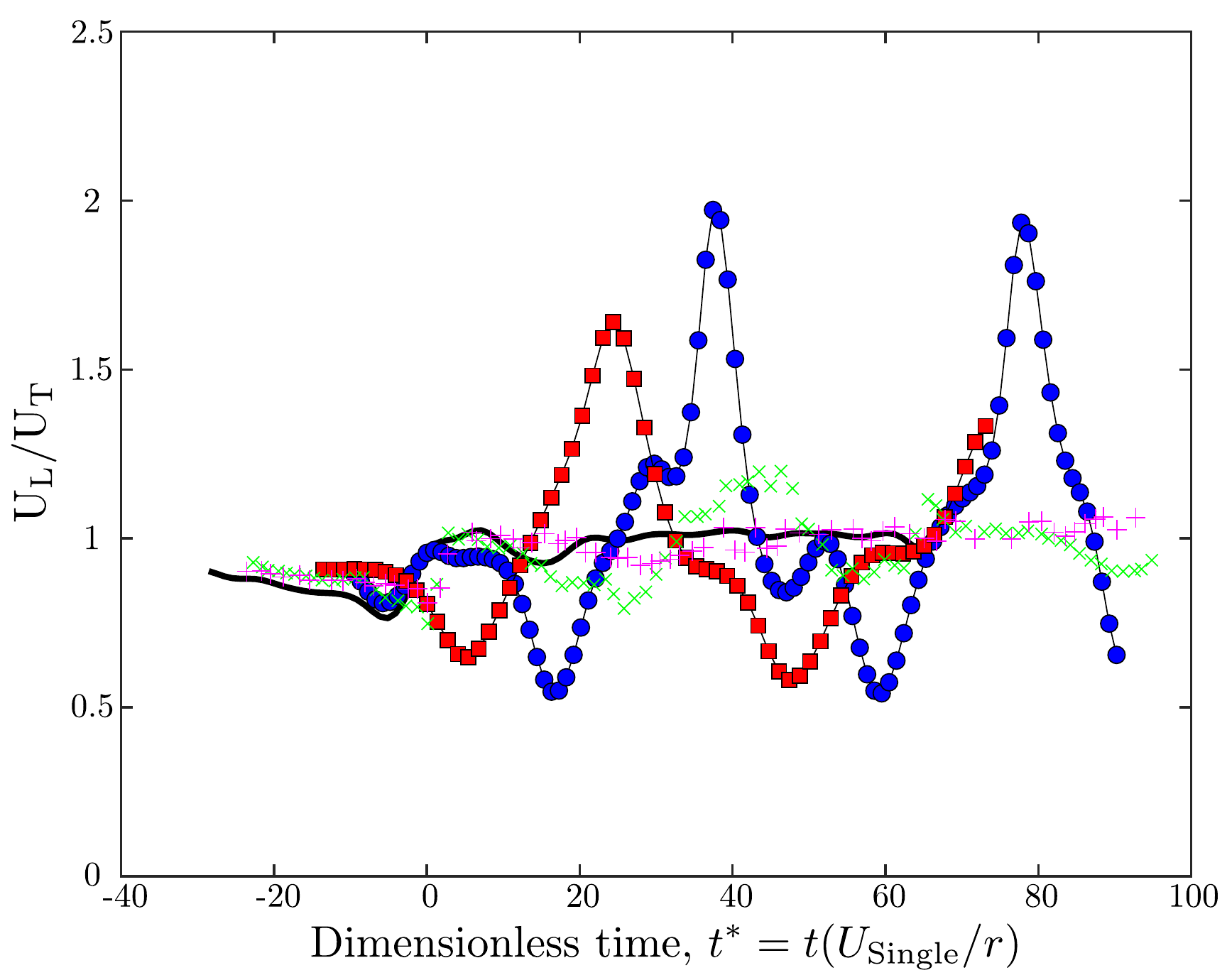}
    \caption{}
    \label{fig:STCDTL}
\end{subfigure}
     \caption{In the shear-thinning inelastic fluid with a power index of n = 0.85 ( \tikzplus[draw=magenta]{10pt}) \& n = 0.55 ( \tikzcross[draw=green]{6pt}): (\ref{fig:STCharacteristic}) Dimensionless distance between the two bubbles as a function of dimensionless time; (\ref{fig:STCDTL}) Velocity ratio of the leading bubble (U$_{\text{L}}$) to the trailing bubble (U$_{\text{T}}$) as a function of dimensionless time.  The results are compared with that of the bubbles in the Newtonian fluid (---), supercritical bubbles in VE1 fluid ( \tikzcircle[fill=blue]{4pt} ), and VE2 fluid ( \tikzsquare[fill=red]{10pt}). The results for the shear-thinning inelastic fluids were reproduced from Velez-Cordero et al. \cite{velez2011hydrodynamic}.}
    \label{fig:ST}
\end{figure}

Velez-Cordero et al. \cite{velez2011hydrodynamic} studied the hydrodynamic interaction between two bubbles rising in the shear-thinning inelastic fluid. They reported that the inline bubbles exhibit drafting and kissing phases similar to the two-bubble systems in Newtonian fluid. Instead of the final tumbling phase, the bubbles switch their relative positions and weakly oscillate about the horizontal axis. Nonetheless, these oscillations eventually dampen out and the bubbles rise as a stable doublet. They concluded these oscillations resulted from non-Newtonian effects. The results for the two-bubble interaction in the shear-thinning inelastic fluid from Velez-Cordero et al. \cite{velez2011hydrodynamic} is compared to the present study. Fig. \ref{fig:STCharacteristic} shows the dimensionless distance between the bubbles as a function of dimensionless time in shear-thinning inelastic fluid with power index, n = 0.85 and n = 0.55. At a comparable Reynolds number of 2 $<$ $Re$ $<$ 6.4, as the strength of the shear-thinning effect increases, the tendency of the bubbles to stay together increases after the kissing phase, thus hindering the bubble pair from tumbling or separating. 

As seen in Fig. \ref{fig:STCDTL}, the velocity ratio of the leading bubble to the trailing bubble in the shear-thinning inelastic fluid varies only about the unit value. Whereas, it overshoots in viscoelastic shear-thinning fluid as the bubbles continue to oscillate about the horizontal axis. This implies that the presence of elasticity and/or a negative wake increases the oscillation experienced by the bubbles. Further, no tumbling phase signifies that the inertia is not sufficiently large. As Zenit and Feng \cite{zenit2018hydrodynamic} mentioned, large bubble deformation and low inertia favor the stable doublet formation. Introducing the elasticity to the fluid, however, may hinder or enhance the DKD interaction depending on its strength compared to the viscous forces. 

\section{Conclusions}

In this work, the hydrodynamic interaction between a pair of bubbles rising inline in the viscoelastic shear-thinning fluids was studied experimentally. Distinct bubble-bubble interaction in the subcritical and supercritical bubble volumes was reported. The results from the experiments could be summarized as:

\begin{enumerate}
  \item For a subcritical bubble pair, the trailing bubble accelerates in the wake of the leading bubble. Hence, the drafting-kissing phases are observed similar to that of the bubbles in Newtonian fluid. However, instead of the final tumbling phase, the bubbles coalesce with each other. 
  
    \item From the PIV results it is suggested that, since the lower stagnation point is far away from the leading bubble, the bubbles coalesce with each other on contact.
  \item For a supercritical bubble pair, the bubbles exhibit a new interaction phenomenon referred to as the drafting-kissing-dancing (DKD). When the trailing bubble is introduced in the wake of the leading bubble, the flow fields around the leading bubble are strongly modified. Because of this, the bubbles adopt a diagonal alignment. The trailing bubble, accelerating in the wake of the leading bubble, actively interchanges its relative position with the leading bubble. This cycle repeats as the bubble-pair rises to the free surface and is referred to as ``dancing". 
  \item Therefore, the dancing phase results from the elasticity, deformability of the bubble surface and the presence of a negative wake. A trade-off between the elastic and viscous forces determines the amplitude of oscillation. Whereas, in the absence of elasticity, the bubbles in shear-thinning inelastic fluids formed a more or less stable doublet pair. 
  \item Furthermore, the tendency of the bubbles to stay close together (to form clusters) increases with the polymer concentration. By modifying the rheological properties of the fluid, it was shown that the subcritical bubbles can also exhibit an interaction similar to the DKD process.
\end{enumerate} This study provides some understanding of the bubble-bubble interaction in viscoelastic shear-thinning fluids, which are relevant in understanding the clustering phenomenon often observed in the non-Newtonian bubbly flows. Further detailed investigations on the bubbly flows in viscoelastic shear-thinning fluids are required to define a set of conditions for the bubble cluster formations. 

\section*{Appendix}

\subsection*{Bubble-bubble interaction in Newtonian fluid}

\begin{figure}[H]
\centering
    \includegraphics[scale=0.7]{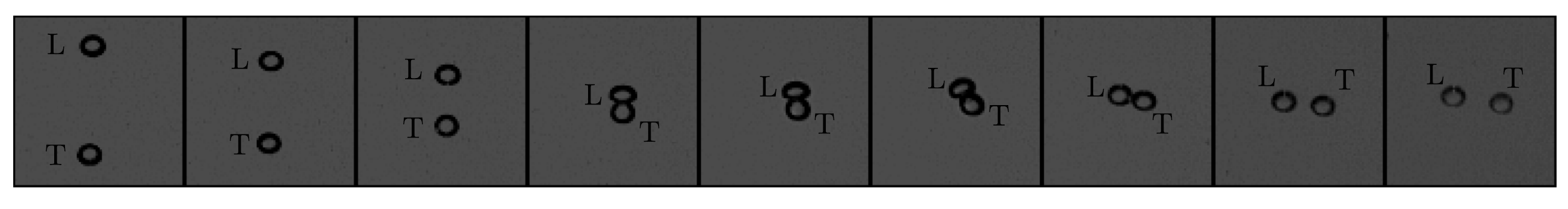}
    \caption{Snapshots of images showing the classical drafting-kissing-tumbling (DKT) process observed between a bubble pair (with a diameter of 5.3 mm) rising inline in the Newtonian fluid at $Re$ $\approx$ 8.}
    \label{fig:Newtonian Interaction}
\end{figure}

Depending on the separation distance and the Reynolds number, the bubbles in the Newtonian fluid can attract or repel each other. When a trailing bubble rises behind the leading bubble, the leading bubble’s wake attracts the trailing bubble. Because of this, the trailing bubble accelerates and catches up with the leading bubble. Thus, the drafting phase occurs. Following that, the bubbles touch each other briefly. After this kissing phase, the bubbles tumble and separate apart. Fig. \ref{fig:Newtonian Interaction} shows the classical drafting-kissing-tumbling (DKT) process observed between the two bubbles rising inline in the Newtonian fluids for a $Re$ $\approx$ 8 \cite{kok1989dynamics,kok1993dynamics,brennen2005fundamentals}.  

\begin{figure}[H]
    \begin{subfigure}[h]{0.51\textwidth}
    \centering
    \includegraphics[scale=0.5]{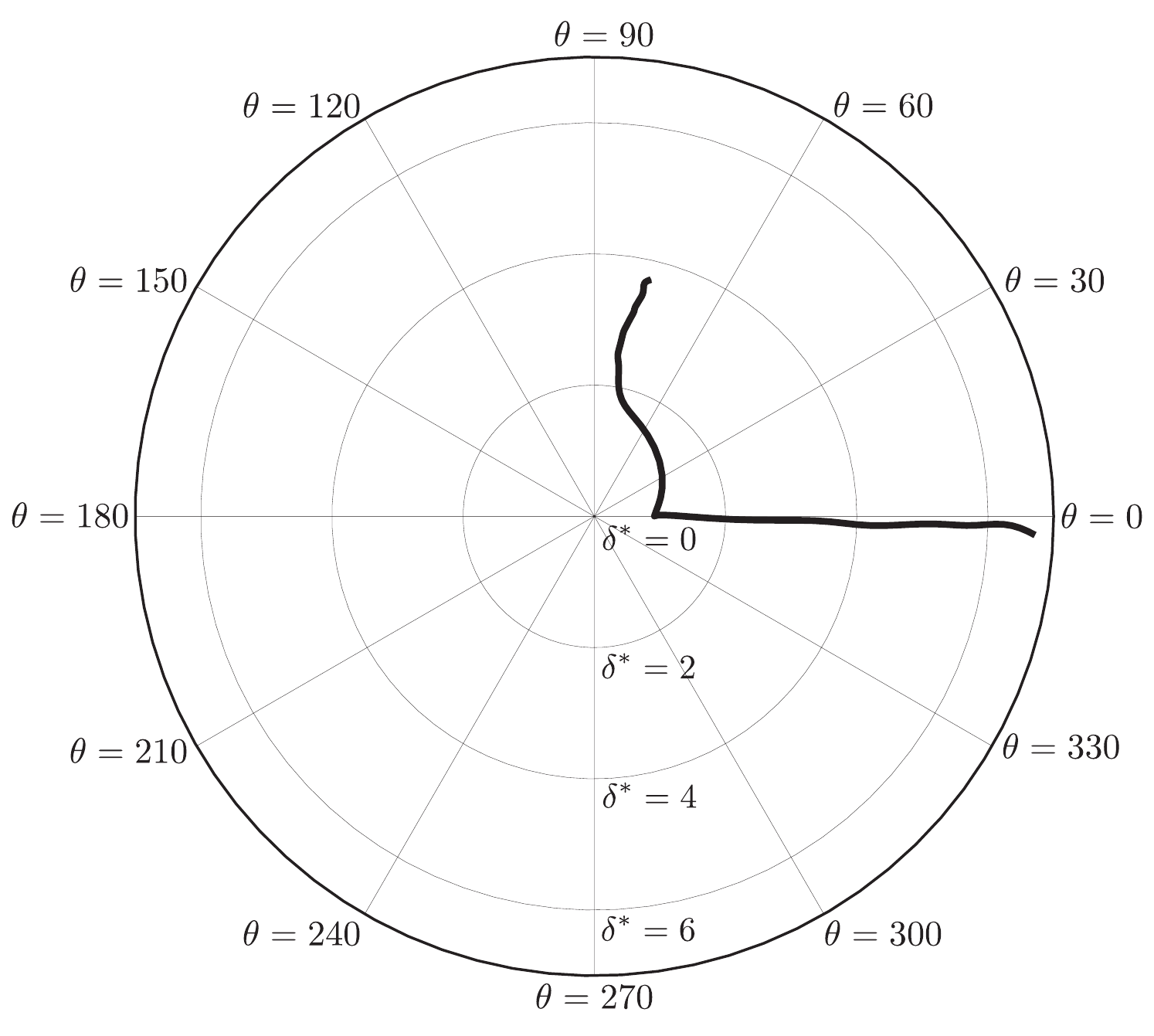}
    \caption{}
    \label{fig:AppendixNewtonianPolar}
\end{subfigure}
\begin{subfigure}[h]{0.51\textwidth}
\centering
    \includegraphics[scale=0.5]{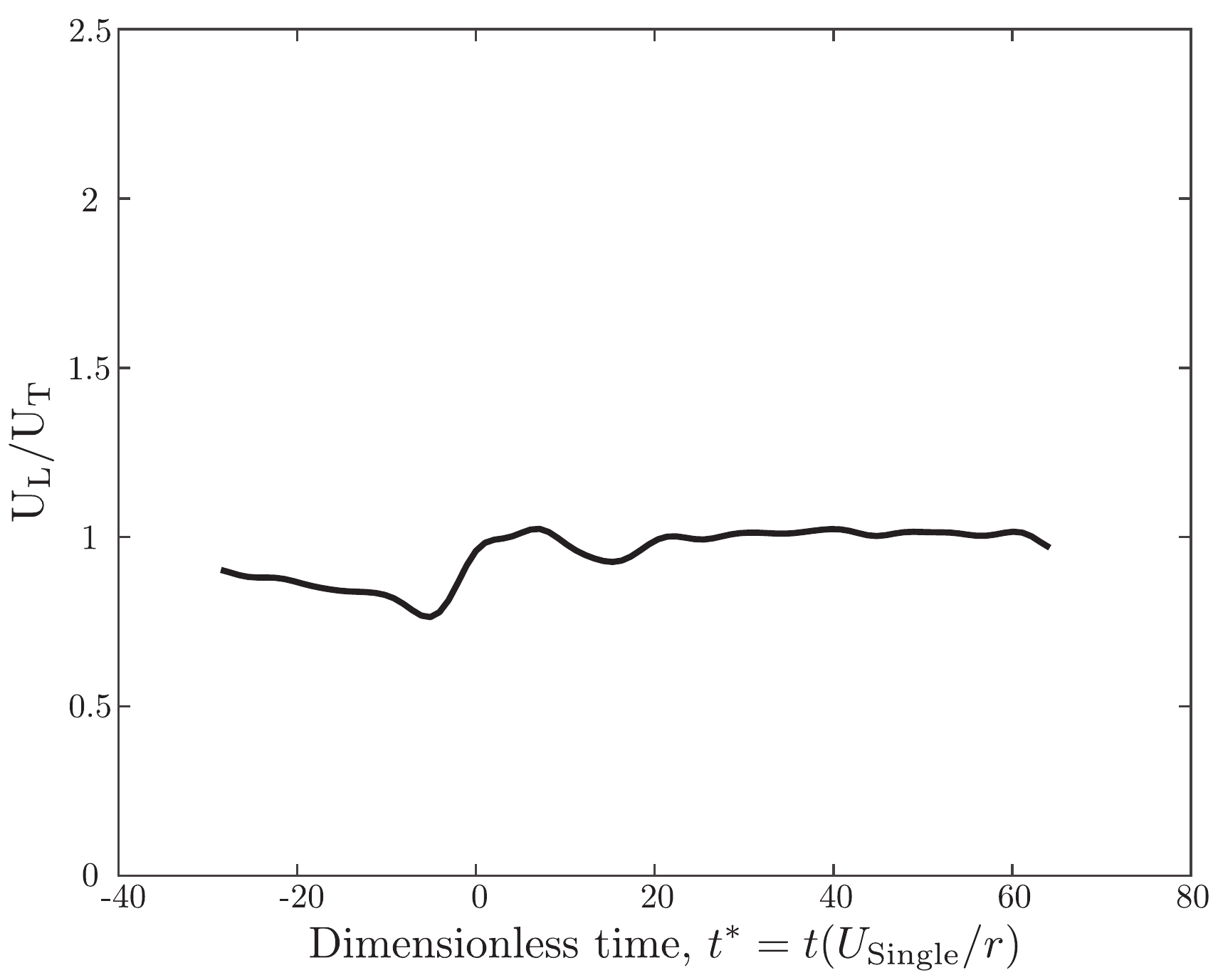}
    \caption{}
    \label{fig:AppendixNewtonianCDTL}
\end{subfigure}
     \caption{In the Newtonian fluid (---): (\ref{fig:AppendixNewtonianPolar}) Dimensionless distance between the two bubbles as a function of angle in polar plots; (\ref{fig:AppendixNewtonianCDTL}) Velocity ratio of the leading bubble (U$_{\text{L}}$) to the trailing bubble (U$_{\text{T}}$) as a function of dimensionless time. }
\end{figure}

Fig. \ref{fig:AppendixNewtonianPolar} shows the dimensionless distance between two bubbles as a function of angle in the Newtonian fluid. During the drafting-kissing phase the angle between the bubbles with the vertical axis remains close to 0\textdegree. After the kissing phase, the angle between the bubbles reaches almost 90\textdegree. Similarly, from the corresponding drag ratio of the trailing bubble to the leading bubble (Fig. \ref{fig:AppendixNewtonianCDTL}), it can be inferred that drag ratio decreases during the drafting phase only to increase in the kissing phase. After the tumbling phase, the drag ratios of the bubble are approximately equal to one. They rise as if they are rising in an isolated fluid. 

\subsection*{Particle-particle interaction in viscoelastic fluid}

\begin{figure}[H]
\centering
    \includegraphics[scale=0.8]{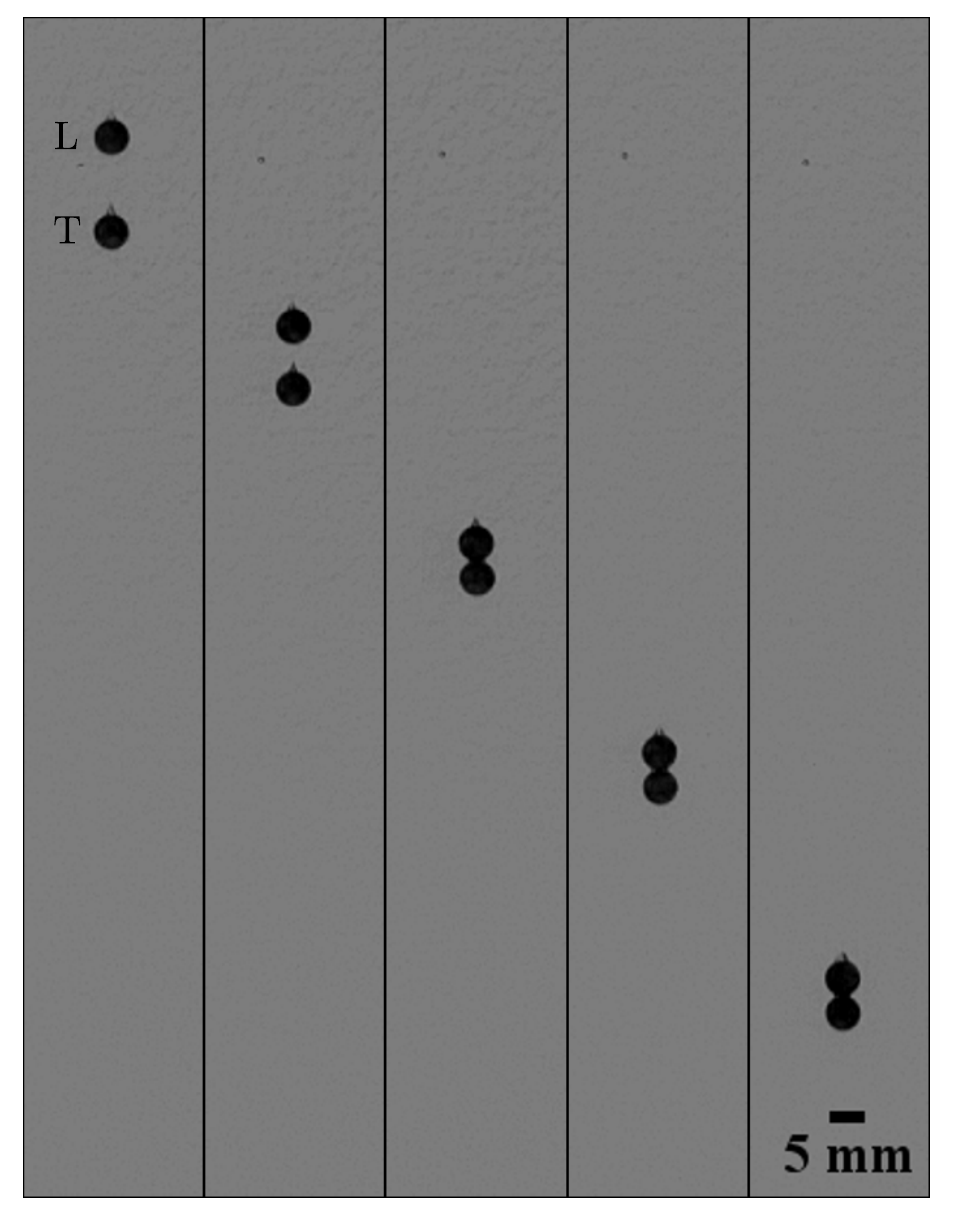}
    \caption{Series of images showing the interaction between two sedimenting solid spheres of diameter 4.75 mm in the VE1 viscoelastic fluid for a  $Re$ $\approx$ 5 and Wi = 440. Interval between the images is 150 millisecond. 
    }
    \label{fig:MontageSpheres}
\end{figure}

\begin{figure}[H]
    \begin{subfigure}[h]{0.51\textwidth}
    \centering
    \includegraphics[scale=0.5]{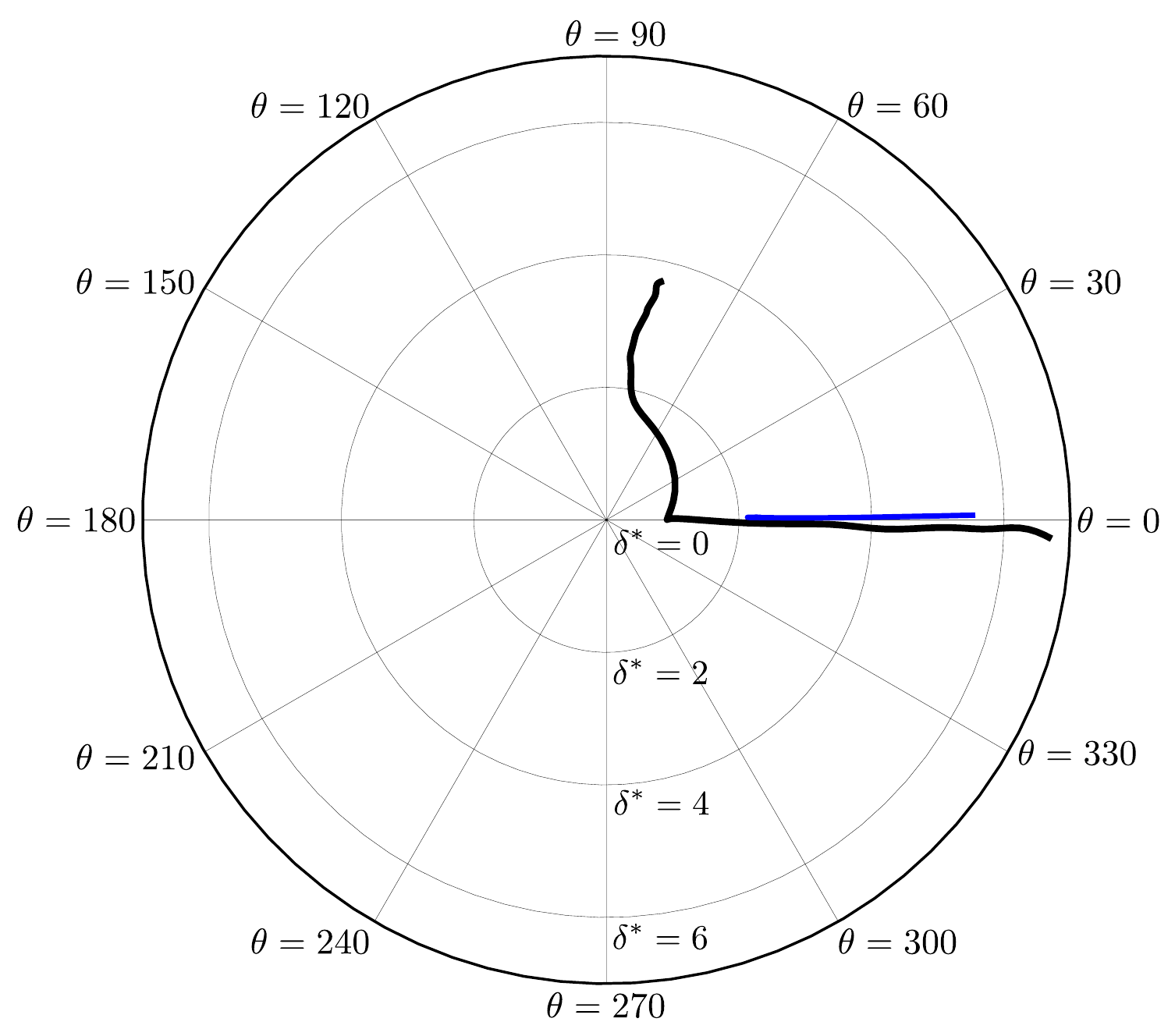}
    \caption{}
    \label{fig:AppendixParticlePolar}
\end{subfigure}
\begin{subfigure}[h]{0.51\textwidth}
\centering
    \includegraphics[scale=0.5]{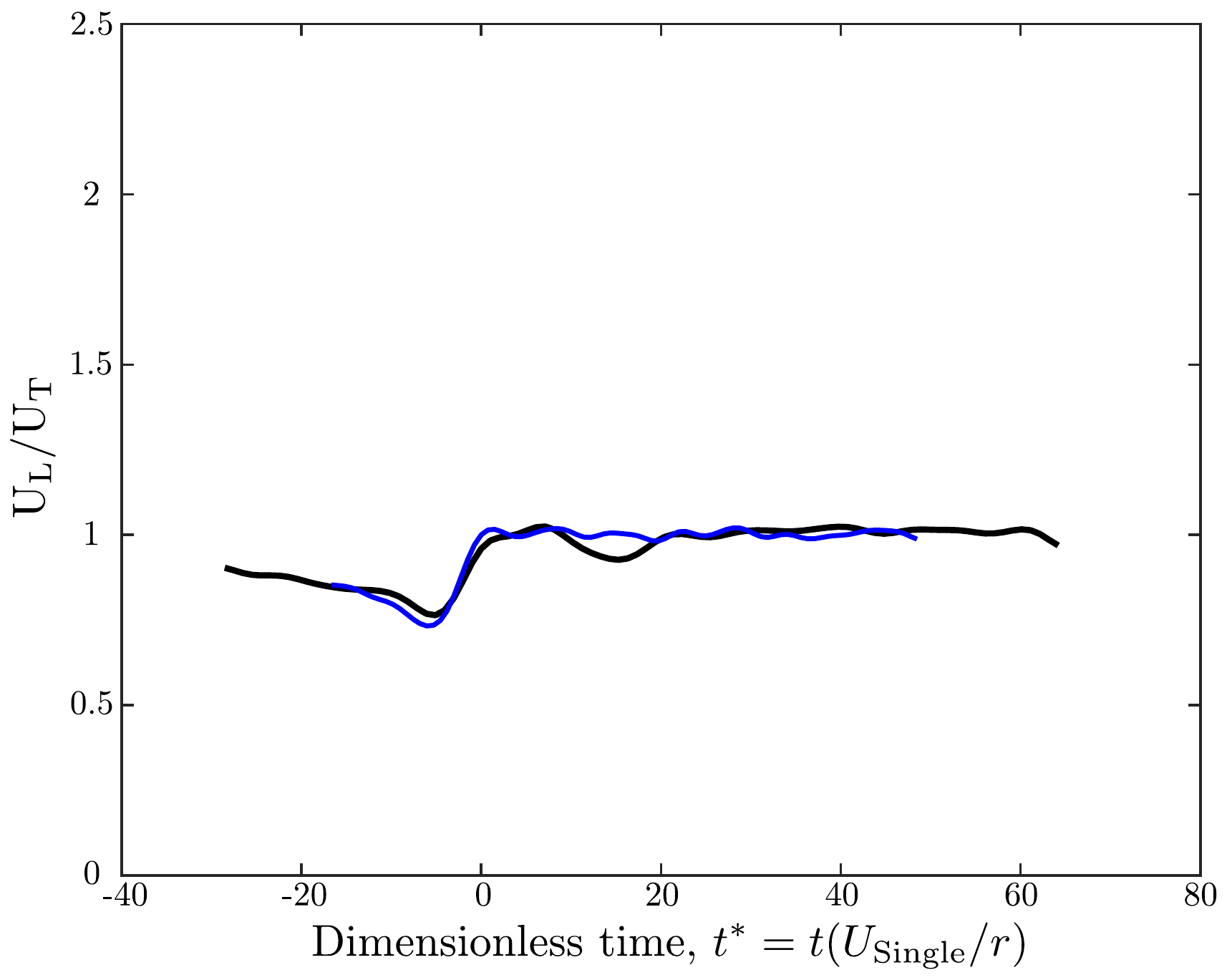}
    \caption{}
    \label{fig:AppendixParticleCDTL}
\end{subfigure}
     \caption{In VE1 fluid (\tikzline[draw=blue]{10pt}): (\ref{fig:AppendixParticlePolar}) Dimensionless distance between the two solid spheres as a function of angle in polar plots; (\ref{fig:AppendixParticleCDTL}) Velocity ratio of the leading sphere (U$_{\text{L}}$) to the trailing sphere (U$_{\text{T}}$) as a function of dimensionless time. The results are compared to that of the classical DKT process observed for the two bubbles in the Newtonian fluid (---). }
\end{figure}

Joseph et al. \cite{joseph1994aggregation} studied the interaction between two solid spheres sedimenting side-by-side in a viscoelastic fluid. They concluded that elasticity, along with the shear-thinning effect, attracts the trailing sphere to the leading sphere's wake. Thus, the trailing sphere accelerates and catches up with the leading sphere. Fig. \ref{fig:MontageSpheres} shows the snapshots of the interaction between two solid spheres settling inline in the VE1 fluid. After the kissing phase, however, no tumbling phase is observed. Instead, the spheres form a stable aggregate \cite{gumulya2011effects,ardekani2008two}. This confirms that the elastic forces are dominant over inertia. From Fig. \ref{fig:AppendixParticlePolar}, it can be observed that the dimensionless distance and the angle between the two solid spheres remains unchanged after the kissing phase. Similarly, from the corresponding drag ratio of the trailing sphere to the leading sphere (Fig. \ref{fig:AppendixParticleCDTL}), it can be inferred that drag ratio remains close to one. Therefore, the particles settle as a vertical pair. This is different for the case of bubbles owing to their deformability. 

\section*{Supplemental Material}

Supplemental video S1: Drafting-kissing-dancing bubble-pair interaction in viscoelastic shear-thinning fluid [\href{https://drive.google.com/file/d/1yC9D8QWQI23fPQtSraSc5SKB2MXqDZpW/view?usp=sharing}{Link}]

\bibliography{references}

\end{document}